\documentclass[11pt]{article}
\usepackage{float}
\usepackage{amsmath}
\usepackage{graphicx}
\usepackage{amssymb}
\usepackage{amsmath, amsthm, amssymb}
\usepackage{geometry}
\usepackage{mathtools}
\usepackage{setspace}
\usepackage{xcolor}
\usepackage{enumerate}
\usepackage{bm}
\usepackage{makecell}
\usepackage{subfloat} 
\usepackage{hyperref}
\usepackage{optidef}
\hypersetup{colorlinks=true, linkcolor = blue}
\usepackage{booktabs}
\usepackage{caption}
\usepackage{subcaption}
\usepackage[font=small,labelfont=bf]{caption}

\usepackage{optidef}
\usepackage{setspace}
\usepackage{enumitem}
\usepackage{lscape}

\usepackage{natbib}
\usepackage[nodots]{numcompress}
\usepackage{fancyhdr}
\graphicspath{ {Images/} }

\newcounter{daggerfootnote}

 \title{{A Stability-Driven Framework for Long-Term Hourly Electricity
Demand Forecasting}}
  
        \vspace{20pt} \author{\normalsize Soumyadeep Dhar\thanks{These authors contributed equally.}, Ayushkumar Parmar\footnotemark[1], \\ \normalsize Haifeng Qiu\thanks{corresponding author}, Juan Ramon L. Senga, S. Viswanathan\thanks{\noindent \emph{Dhar:} Indian Institute of Technology Kharagpur, Kharagpur 721302, India. \protect\href{mailto:sdhar1602@kgpian.iitkgp.ac.in}{{sdhar1602@kgpian.iitkgp.ac.in}}. \emph{Parmar:}  Indian Institute of Technology Kharagpur, Kharagpur 721302, India. \href{mailto:ayushparmar_ise@kgpian.iitkgp.ac.in}{{ayushparmar\textunderscore ise@kgpian.iitkgp.ac.in}}. \emph{Qiu:} Energy Research Institute @ NTU (ERI@N), Singapore 637141. \protect\href{mailto:haifeng.qiu@ntu.edu.sg}{{haifeng.qiu@ntu.edu.sg}}. \emph{Senga:} Center for Energy and Environmental Policy Research, Massachusetts Institute of Technology, Cambridge, MA, USA. MIT Climate Policy Center, Massachusetts Institute of Technology, Cambridge, MA, USA. \protect\href{mailto:jsenga@mit.edu}{{jsenga@mit.edu}}. \emph{Viswanathan:} Energy Research Institute @ NTU (ERI@N), Singapore 637141. Nanyang Business School, Nanyang Technological University, Singapore 639798. \protect\href{mailto:asviswa@ntu.edu.sg}{{asviswa@ntu.edu.sg}}.
}} 

\vspace{10pt}
\begin{document}
\thispagestyle{empty}
\maketitle
\date{}
\onehalfspacing
\vspace{-30 pt}
\begin{abstract}
    \noindent Long-term electricity demand forecasting is necessary for grid and operations planning, as well as for the analysis and planning of energy transition strategies. However, accurate long-term load forecasting with high temporal resolution remains a challenge as most existing approaches focus on aggregated forecasts, which require accurate prediction of many other variables as bottom-up sectoral forecasts are developed and aggregated. In this study, we propose a parsimonious methodology that uses t-tests to verify load stability and the correlation of load with gross domestic product (GDP) to arrive at a long-term hourly load forecast.  Applying this method to Singapore’s electricity demand, analysis of multi-year historical data (2004–2022) reveals that its relative hourly load has remained statistically stable, with an overall percentage deviation of 4.24\% across seasonality indices. Utilizing these stability findings, five-year-ahead total yearly forecasts were generated using gross domestic product (GDP) as a predictor, and the hourly loads were then forecasted using hourly seasonality index fractions. The maximum Mean Absolute Percentage Error (MAPE) across multiple experiments of six years ahead forecasts was only 6.87\%. We then further apply the methodology to Belgium (an OECD country) and Bulgaria (a non-OECD country), to arrive at MAPE values of 6.81\% and 5.64\%, respectively. Additionally, stability results were incorporated into a short-term forecasting model based on exponential smoothing, demonstrating comparable or improved accuracy relative to existing machine learning-based methods. The findings show that parsimonious approaches can be effective in arriving at long-term, high resolution forecasts.
    

\end{abstract}

\thispagestyle{empty}

\clearpage

\onehalfspacing
\setcounter{footnote}{0}
\renewcommand{\thefootnote}{\arabic{footnote}}
\setcounter{page}{1}

    \section{Introduction}
	
\begin{figure}[!h]	
  \centering
			\includegraphics[width=0.9\textwidth]{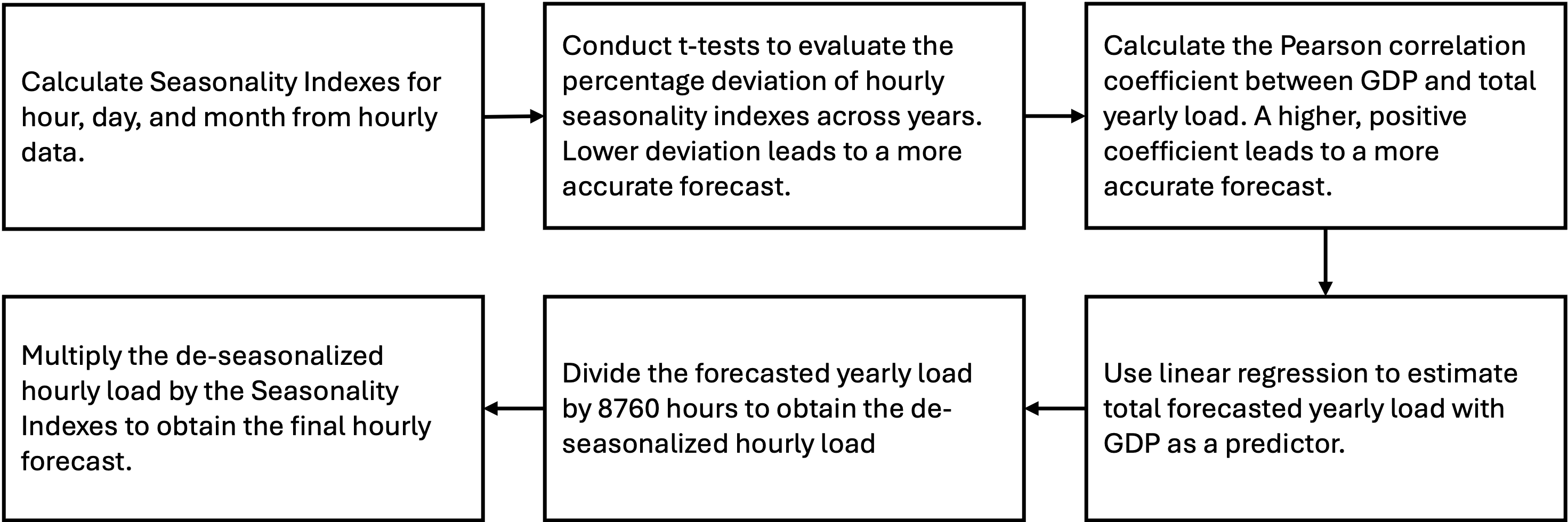}
        \caption{\textbf{Proposed Methodology for Long-term Load Forecasting with Hourly Resolution.}} 
        \label{Fig_MethodChart}
\end{figure}

Long-term electricity demand forecasting is necessary for planning future capacity investments in generation and transmission, managing electricity markets and the grid, and identifying  decarbonization pathways (\citeauthor{ho_regional_2021}, \citeyear{ho_regional_2021}; \citeauthor{van_ouwerkerk_impacts_2022}, \citeyear{van_ouwerkerk_impacts_2022}). In this study, we introduce a new long-term electricity load forecasting methodology that uses the stability of load and the correlation of load with Gross Domestic Product (GDP) to arrive at a long-term, high-resolution hourly forecast. Most long-term forecasts are built through bottom-up approaches that focus on the growth of sectors and the technological advancement of specific electrification drivers such as electric vehicles or heat pumps (\citeauthor{oh_forecasting_2016}, \citeyear{oh_forecasting_2016}; \citeauthor{massier_electrification_2018} \citeyear{massier_electrification_2018}).  Predicting future usage patterns of these new technologies is used to provide a representation of future demand and consequently the hourly demand forecast. Long-term forecasts also often look at aggregated yearly or monthly load instead of hourly demand which is what would be needed in more complex power systems analysis (\citeauthor{hamed_forecasting_2022}, \citeyear{hamed_forecasting_2022}; \citeauthor{kazemzadeh_hybrid_2020}, \citeyear{kazemzadeh_hybrid_2020}). We deviate from these approaches and propose a methodology that gives a long-term forecast at an hourly resolution while using top-down indicators. This provides a parsimonious alternative to traditional bottom-up methods which relies on multiple assumptions that may be difficult to verify. Our methodology instead relies on establishing only two, statistically verifiable assumptions: first, that historical hourly load is \textit{Stable} and second, that annual load is correlated with GDP. We then use linear regression on total annual load with GDP as a predictor. Finally, the stability results are used to translate the yearly predicted load to an hourly load. Fig. \ref{Fig_MethodChart} summarizes the methodology. Our methodology is considerably simpler than the few studies that have looked at long-term hourly resolution forecasts. For example, \citeauthor{gonzalez_grandon_electricity_2024} (\citeyear{gonzalez_grandon_electricity_2024}) uses Long Short-Term Memory (LSTM) machine learning models while \citeauthor{royal_statistical_2025} (\citeyear{royal_statistical_2025}) uses hybrid statistical models that account for weather data and district occupancy trends to arrive at the resolution. However, we show that our approach still yields low error rates.

According to \citeauthor{nti_electricity_2020} (\citeyear{nti_electricity_2020}), electricity forecasting can be subdivided into (a) three main methodologies: correlation, extrapolation, or a combination of both, and (b) four forecast horizons: (i) very short-term (real time control), (ii) short-term (1 hour to 7 days), (iii) medium-term (7 days to 1 year), and (iv) long-term (1 year to 20 years). The methodology devised in this study is an example of a long-term forecast that is a combination of extrapolation and correlation since we use both GDP to correlate total yearly demand, and use statistical analysis through seasonality indices to arrive at our final hourly forecast.




Load stability is defined as how much the load follows the same predictable patterns across various time horizons (i.e., daily, weekly, yearly). Literature on load stability remains scarce but incorporating seasonal patterns in load forecasts has been shown to improve accuracy. For example, it was demonstrated that pattern-based statistical methods could effectively account for seasonal fluctuations and medium-term trends in energy consumption (\citeauthor{en16020827}, \citeyear{en16020827}) and that incorporating seasonal climate forecasts can significantly improve medium-term electricity demand forecasting accuracy (\citeauthor{DEFELICE2015435}, \citeyear{DEFELICE2015435}). Borrowing terminology and concepts from exponential smoothing literature, we use seasonality indices (SIs) -- an indicator of how much demand deviates from the average in each \textit{season} -- and the difference between SIs across years to calculate how much hourly SIs changes from year to year. Small changes correspond to more stable load while larger changes signify less stable load. 


We demonstrate our forecasting methodology on a case study of the Republic of Singapore. Singapore serves as an ideal test application because its small size relative to other countries reduces other confounding factors that affect load. Previous attempts at load forecasting for Singapore have either only looked at long-term bottom up approaches incorporating the residential (\citeauthor{chuan_modeling_2015}, \citeyear{chuan_modeling_2015}), transportation (\citeauthor{massier_electrification_2018}, \citeyear{massier_electrification_2018}), and cooling (\citeauthor{oh_forecasting_2016}, \citeyear{oh_forecasting_2016}) sectors. Meanwhile, seasonality and load stability has only been applied in short-term forecasts such as when \citeauthor{huang_electricity_2017} (\citeyear{huang_electricity_2017}) used a Holt-Winters model with four-fold seasonality to show that it performs better than a triple seasonality model in forecasting Singapore's next half-hourly demand or when \citeauthor{deng_short-term_2010} (\citeyear{deng_short-term_2010}) showed that the multiplicative seasonal model outperforms the ARIMA model in short-term forecasts. We use a series of t-tests to demonstrate that the overall percentage deviation from the mean of seasonality indices is 4.24\% from 2004 to 2022. This shows that Singapore’s hourly electricity demand coefficient adjusted for seasonality has remained stable throughout the duration. At the same time, Singapore's electricity demand and GDP are highly correlated with a Pearson correlation coefficient of 0.9947. Using GDP as a predictor for total load and dividing this accordingly based on the calculated seasonality indices leads to our forecast. The results show Mean Absolute Percentage Errors (MAPE) of 6.87\% or lower for five year-ahead hourly forecasts across six forecast years. The methodology is then extended to two other countries: Belgium and Bulgaria, representing an Organization for Economic Cooperation and Development (OECD) and Non-OECD country respectively. The MAPE values are 6.81\% and 5.64\%, respectively for two-year ahead forecasts.

\begin{figure}[!h]	
  \centering
			\includegraphics[width=0.8\textwidth]{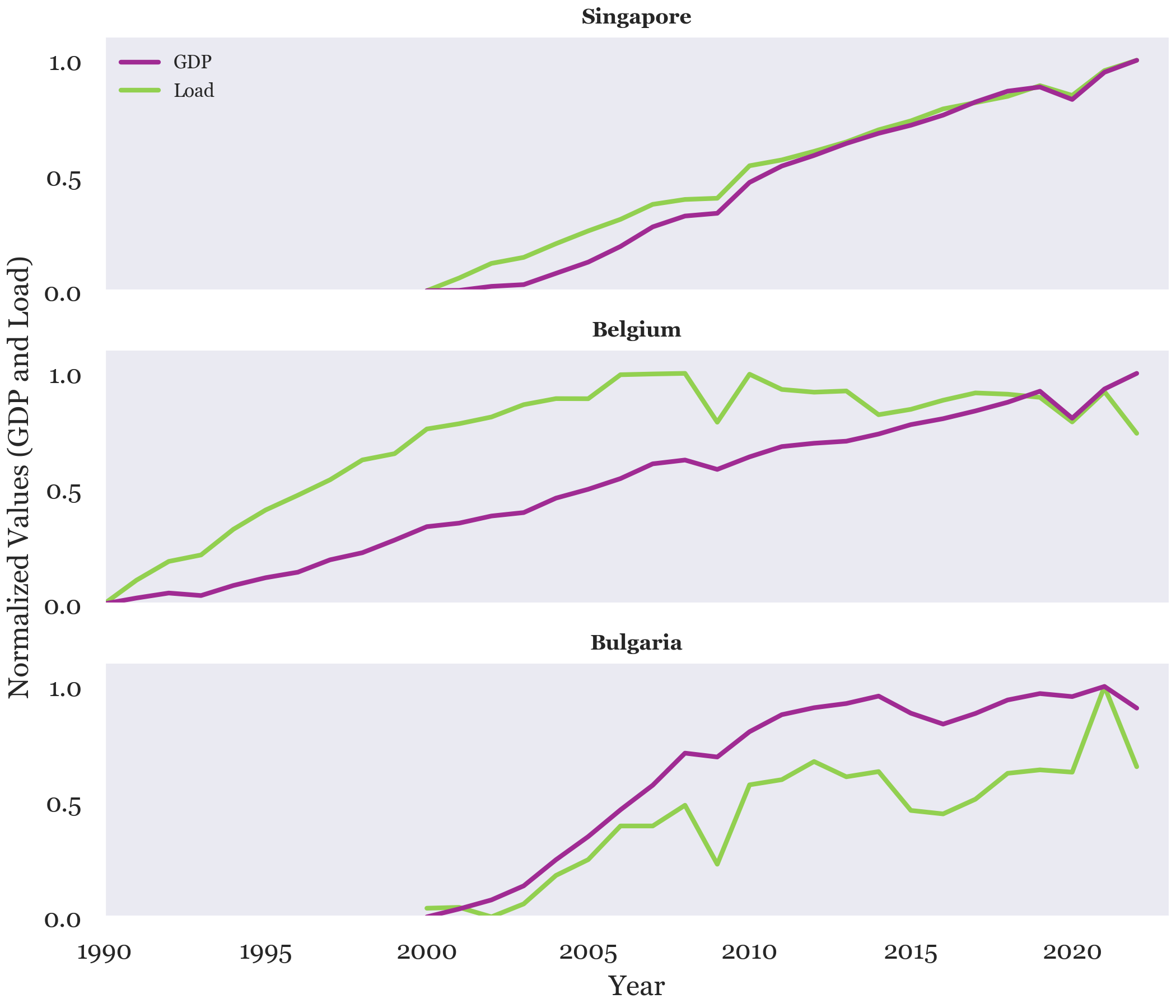}
			\label{GDP_and_load_correlation}
		\hfill
        \caption{\textbf{Normalized GDP and Load values for (a) Singapore, (b) Belgium, and (c) Bulgaria (\citeauthor{owid}, \citeyear{owid}).} The Pearson correlation coefficient for each country are: Singapore: 0.9947; Belgium: 0.8084; Bulgaria: 0.9204.} 
        \label{Figure_GDPLoadCorr}
\end{figure}

We use GDP -- a measure of the total monetary value of all goods and services produced within a country -- as the predictor since it has been shown to correlate with electricity consumption even when there have been efficiency gains in electricity usage (\citeauthor{Steinbuks2017}, \citeyear{Steinbuks2017}; \citeauthor{su12155931}, \citeyear{su12155931}). For example, a 2024 panel analysis of 31 countries in Latin America and the Caribbean showed that a 1\% increase in GDP leads to about a 1.5\% rise in electricity demand (\citeauthor{BAZANNAVARRO2024e33521}, \citeyear{BAZANNAVARRO2024e33521}).
Similarly, country-specific econometric models often show high explanatory power when using GDP as the key predictor. In Singapore's case, it was shown that there is a causal relationship between economic growth and electricity consumption (\citeauthor{sharif_electricity_2017}, \citeyear{sharif_electricity_2017}). Fig. \ref{Figure_GDPLoadCorr} shows the normalized GDP and electricity consumption for Singapore, Belgium, and Bulgaria. Past literature has shown that incorporating other econometric factors such as population, renewable share, and real price of certain goods improves annual forecasts (\citeauthor{hamed_forecasting_2022}, \citeyear{hamed_forecasting_2022}). However, we are able to show that GDP can provide enough explanatory power to arrive at relatively accurate forecasts.

The primary contribution of this research lies in its integration of statistical demand stability analysis with a top-down, GDP-based long-term forecasting framework. Existing studies have extensively explored short-term electricity demand prediction using regression methodologies, machine learning algorithms (\citeauthor{konica_forecasting_2016}, \citeyear{konica_forecasting_2016}; \citeauthor{zahid_electricity_2019}, \citeyear{zahid_electricity_2019}; \citeauthor{bouktif_optimal_2018}, \citeyear{bouktif_optimal_2018}; \citeauthor{neo_hybrid_2023}, \citeyear{neo_hybrid_2023}; \citeauthor{quan_short-term_2014}, \citeyear{quan_short-term_2014}) and long-term forecasting of aggregated load data (\citeauthor{en16020827}, \citeyear{en16020827}; \citeauthor{10.11648/j.ijepe.20241302.11}, \citeyear{10.11648/j.ijepe.20241302.11}). However,  research that rigorously validates and uses hourly-level demand pattern stability remains scarce. This gap is critical because stable and predictable hourly demand structures enable more cost-effective infrastructure investments, enhanced renewable energy integration, and informed policy making for the future. This study also addresses a key limitation of prior forecasting models, which are often region-specific and lack cross-country validation. The ability to forecast long-term hourly electricity demand across diverse economies highlights the applicability of the approach for energy planning in both developed and industrializing nations.

\section{Stability of Load and Seasonality indices}\label{Section_Seasonality_Index}

\subsection{Data Description}\label{Section_Data}

To build our hypothesis and the methodology we have developed, we first establish the test-case for Singapore, then we extend it for 2 European countries: Belgium(an OECD country) and Bulgaria(a non-OECD country). We first use half-hourly system-wide electricity demand data (in MW) for Singapore from Jan 1, 2004 to Dec 31, 2022 obtained from Singapore's Energy Market Authority (\citeauthor{ema_half-hourly_2023}, \citeyear{ema_half-hourly_2023}). Starting January 1, 2014, the data includes separate tracking of (1) system-wide and (2) National Electricity Market (NEM) demand data. NEM demand is the actual demand on all Singapore generation registered facilities (e.g. powerplants). Meanwhile, system-wide demand is the NEM demand plus the demand satisfied by consumers with their own embedded generators (e.g. rooftop solar). Furthermore, January 1, 2014 also marks the point where the half-hourly EMA forecasts are also included in the data set. In our analysis, we transform the half-hourly data into its hourly equivalents by adding together the half-hourly demands, starting from 00:30 to 01:00 to get the \textit{total demand} for 01:00, adding 01:30 and 02:00 to get \textit{total demand} for 02:00, etc. Note that this is not the MWh electricity consumption. To get consumption per hour, this value would need to be divided by two.\footnote{In transforming half-hourly demand data (MW) to hourly equivalents, we sum consecutive half-hourly values to construct an hourly demand metric assigned to the hour ending at time \textit{T}. This sum represents the aggregate of average power values but does not equate to energy consumption. To derive actual energy (MWh), the hourly demand metric is divided by 2 reflecting the 0.5-hour duration of each constituent interval. As an example, 5600MW consumed from 12:01-12:30 and 5800MW consumed for 12:31-13:00 is tagged as 11,400MW in our model and is equivalent to 5700MWh for 12:01-13:00. The MWh equivalent can be understood as follows: the system draws 5600MW for 30 minutes which is equivalent 2800MWh and 5800 for another 30 minutes equivalent to 2900MWh. For the entire hour, the power consumed is 2800MWh + 2900MWh = 5700MWh.} The transformation is for ease of comparison between forecast and actual demand and to retain the format of MW so that we remain consistent with the half-hourly data that the EMA provides.  This facilitates easier calculations while not removing the daily variations in electricity demand. Where it will be needed for exposition of the analysis, we will convert to MWh. Otherwise, we retain the MW measure. In this paper, we also use electricity \textit{load} and \textit{demand} interchangeably.

\subsection{Stability of Hour of Day Seasonality Indices of Singapore's Electricity Demand}\label{Section_Stable}

\begin{figure}[!h]
  \centering
			\includegraphics[width=\textwidth]{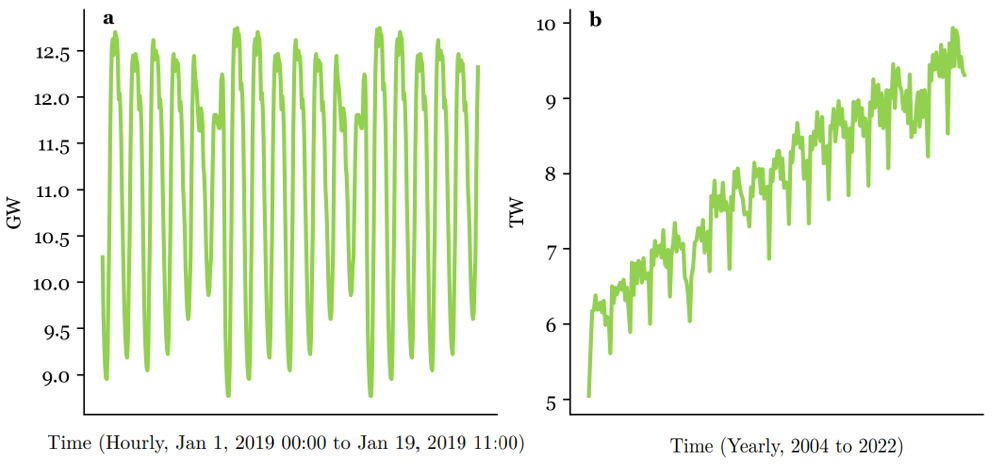}
        		\caption{\textbf{Historical hourly (a) and monthly (b) load for Singapore.} The hourly load in (a) is from January 1, 2019 00:00 to January 19, 2019 11:00. The monthly load is from 2004 to 2022.}
                \label{Fig_HistoricalLoad_SG}
\end{figure}

In this section, we show that the electricity demand in Singapore has remained stable from 2004-2022. While it has been regularly referenced that Singapore enjoys stable electricity demand (\citeauthor{doi:10.1073/pnas.2026596118}, \citeyear{doi:10.1073/pnas.2026596118}; \citeauthor{LOI2016735}, \citeyear{LOI2016735}; \citeauthor{LOI2018415}, \citeyear{LOI2018415}), there is no rigorous definition of how stable it is. This section attempts to establish that rigor. 
First, based on a visual examination of the data, we observe that Singapore's electricity demand exhibits hourly, daily, and monthly calendar patterns (Fig. \ref{Fig_HistoricalLoad_SG}a and \ref{Fig_HistoricalLoad_SG}b). This observation led to our primary use of seasonality indices as a means to calculate stability. Our goal is to provide a seasonality index (SI) for every hour of each day from 2004 to 2022 and show that across all years, this seasonality index has remained relatively stable. 

We sequentially obtain seasonality indices starting with the month ($SI^1_{y, m}$), day of the week ($SI^2_{y, m, d}$) and hour of the day ($SI^3_{y, m, d, h}$) where $y, m, d, h$ refer to the year, month, day of the week, and hours, respectively. The superscripts $1, 2,$ and $3$ are used to refer to $SI$s for the month, day of the week, and hour, respectively. All results including corresponding scripts can be accessed in the supplementary material.

\subsubsection{SI for the Month}

Let $D_y^{i, h}$ be the electricity demand in hour $h$ of day $i$ in year $y$, where $h = 1, 2, \dots, 24$,  $i = 1, 2, \dots 365$, and $y= 2004, 2005, \dots, 2022$. The total demand for the month is 
\begin{equation}
    D_{y, m} = \sum_{i \in M_{y, m}}\sum_{h = 1}^{24} D_y^{i, h} 
\end{equation}
where $M_{y, m}$ is the set of days that are part of month $m$ (e.g. $M_{y, 1} = \{ 1, 2, \dots, 31\}$ is the days of January) and $|M_{y,m}|$ is the number of days in month $m$ of year $y$. The specification of a year $y$ only matters when $m = 2$, since there are leap years. In general, $|X|$ for a set $X$ is the number of elements in the set. The total demand for the year is then
\begin{equation}
    \mathcal{D}_y = \sum_{m=1}^{12} D_{y, m}
\end{equation}
and the monthly seasonality index for month $m$ and year $y$, $SI_{y, m}^1$ is
\begin{equation}
    SI_{y, m}^1 = \frac{\frac{D_{y, m}}{|M_{y, m}|}}{\frac{\mathcal{D}_y}{365}} = \frac{365 D_{y, m} }{|M_{y, m}|\mathcal{D}_y} = \frac{\text{Average daily demand for month $m$ in year $y$}}{\text{Average daily demand for year $y$}}
\end{equation}
The overall seasonality index for month $m$ across all years $\bar{SI}_m^1$ is the average of $SI_{y, m}^1$ across all years for $m$:
\begin{equation}
    \bar{SI}_m^1 = \frac{\sum_{y = 2004}^{2022}SI_{y,m}^1}{2022-2004+1} = \frac{\text{Sum of $SI$ for month $m$ across all years}}{\text{Number of years}}
\end{equation}

\subsubsection{SI for the Day of the Week}

Let $d = 1, 2, \dots, 7$ correspond to days of the week where $d = 1$ is a Monday, $d = 2$ is a Tuesday, etc. $SI_{y,m,d}^2$ is the seasonality index for day of the week $d$ in month $m$ of year $y$ and is calculated as follows:
\begin{equation}
    SI_{y, m, d}^2 = \frac{\frac{\sum_{i \in W_{y, m, d}}\sum_{h = 1}^{24}D_y^{i, h}}{|W_{y, m, d}|}}{\frac{D_{y, m}}{|M_{y, m}|} \bar{SI}_m^1} = \frac{\text{Average daily demand for day $d$ of month $m$ in year $y$}}{\text{(Average daily demand for month $m$ in year $y$)(Overall $SI$ of month $m$)}}
\end{equation}
where $W_{y, m, d}$ is the set of days that are part of year $y$, month $m$, and week day $d$. We can calculate the overall seasonality index for the day of the week as:
\begin{equation}
    \bar{SI}_{d}^2 = \frac{\sum_{y= 2004}^{2022} \sum_{m = 1}^{12}SI_{y, m, d}^2}{(2022 - 2004 +1)(12)} = \frac{\text{Sum of $SI$ for day of the week $d$ across all months and years}}{\text{Number of month and year combinations}}
\end{equation}

\subsubsection{SI for Hour of the Day}

Finally, we can calculate the seasonality index for each hour of each day of the week as follows:
\small
\begin{equation}
    SI_{y, m, d, h}^3 = \frac{\frac{\sum_{i \in W_{y, m, d}}D_{y}^{i, h}}{|W_{y, m, d}|}}{\frac{\sum_{i \in W_{y, m, d}}\sum_{h = 1}^{24}D_y^{i, h}}{24(|W_{y, m, d}|) (\bar{SI}_m^1 \bar{SI}_d^2)}} = \frac{\text{Average hourly demand for hour $h$ in day of the week $d$ of month $m$ in year $y$}}{(\text{Average hourly demand for day $d$})(\text{Overall monthly $SI$})(\text{Overall weekly $SI$})}
\end{equation}

\begin{equation}
    \bar{SI}_h^3 = \frac{\sum_{y = 2004}^{2022}\sum_{m = 1}^{12}\sum_{d = 1}^7 SI_{y, m, d, h}^3}{(2022 - 2004 + 1)(12)(7)} = \frac{\text{Sum of $SI$ for hour of the day across all days of the week, months, and years}}{\text{Number of month, year, day of the week combinations}}
\end{equation}
\normalsize

Following this methodology, we can obtain Table \ref{Table_MeanSI} which shows the mean SI $\mu^{d, h}$ from 2004 to 2022 for every day of the week $d$ and hour $h$. That is, 
\begin{equation}
    \mu^{d, h} = \frac{\sum_{y = 2004}^{2022} \sum_{m = 1}^{12}SI_{y,m,d,h}^3}{(2022-2004+1)(12)}
    \label{Formula_OverallSeasonalityIndex}
\end{equation}


\begin{table}[h]
\centering
\caption{Mean SI for Each Hour and Day from 2004–2022 ($\mu^{d, h}$)}
\label{Table_MeanSI}
\resizebox{\textwidth}{!}{%
\begin{tabular}{|c|c|c|c|c|c|c|c|}
\hline
\textbf{Time} & \textbf{Monday} & \textbf{Tuesday} & \textbf{Wednesday} & \textbf{Thursday} & \textbf{Friday} & \textbf{Saturday} & \textbf{Sunday} \\
\hline
\textbf{1:00}  & 0.849 & 0.867 & 0.868 & 0.872 & 0.875 & 0.946 & 0.967 \\
\textbf{2:00}  & 0.819 & 0.836 & 0.836 & 0.840 & 0.844 & 0.911 & 0.933 \\
\textbf{3:00}  & 0.800 & 0.816 & 0.816 & 0.820 & 0.823 & 0.886 & 0.908 \\
\textbf{4:00}  & 0.789 & 0.804 & 0.804 & 0.807 & 0.810 & 0.869 & 0.891 \\
\textbf{5:00}  & 0.786 & 0.801 & 0.801 & 0.803 & 0.807 & 0.862 & 0.883 \\
\textbf{6:00}  & 0.807 & 0.820 & 0.819 & 0.821 & 0.824 & 0.869 & 0.886 \\
\textbf{7:00}  & 0.863 & 0.868 & 0.869 & 0.870 & 0.873 & 0.894 & 0.905 \\
\textbf{8:00}  & 0.924 & 0.923 & 0.925 & 0.925 & 0.926 & 0.921 & 0.911 \\
\textbf{9:00}  & 1.013 & 1.014 & 1.017 & 1.016 & 1.012 & 0.993 & 0.952 \\
\textbf{10:00} & 1.067 & 1.067 & 1.069 & 1.069 & 1.065 & 1.045 & 0.992 \\
\textbf{11:00} & 1.097 & 1.095 & 1.095 & 1.096 & 1.093 & 1.078 & 1.022 \\
\textbf{12:00} & 1.106 & 1.103 & 1.102 & 1.103 & 1.102 & 1.086 & 1.032 \\
\textbf{13:00} & 1.094 & 1.089 & 1.087 & 1.090 & 1.090 & 1.071 & 1.026 \\
\textbf{14:00} & 1.103 & 1.099 & 1.098 & 1.101 & 1.100 & 1.060 & 1.030 \\
\textbf{15:00} & 1.108 & 1.104 & 1.103 & 1.107 & 1.106 & 1.053 & 1.032 \\
\textbf{16:00} & 1.104 & 1.100 & 1.100 & 1.103 & 1.101 & 1.041 & 1.029 \\
\textbf{17:00} & 1.103 & 1.099 & 1.097 & 1.100 & 1.096 & 1.030 & 1.024 \\
\textbf{18:00} & 1.087 & 1.082 & 1.080 & 1.083 & 1.076 & 1.015 & 1.020 \\
\textbf{19:00} & 1.055 & 1.049 & 1.048 & 1.049 & 1.043 & 1.011 & 1.026 \\
\textbf{20:00} & 1.058 & 1.052 & 1.052 & 1.052 & 1.043 & 1.035 & 1.062 \\
\textbf{21:00} & 1.046 & 1.041 & 1.040 & 1.040 & 1.029 & 1.036 & 1.074 \\
\textbf{22:00} & 1.023 & 1.017 & 1.018 & 1.017 & 1.008 & 1.030 & 1.074 \\
\textbf{23:00} & 0.973 & 0.968 & 0.969 & 0.970 & 0.966 & 0.998 & 1.038 \\
\textbf{0:00}  & 0.924 & 0.920 & 0.920 & 0.922 & 0.926 & 0.965 & 0.997 \\
\hline
\end{tabular}%
}
\end{table}

\begin{figure}[!h]
		
		\begin{subfigure}[b]{0.4\textwidth}
			\includegraphics[width=\textwidth]{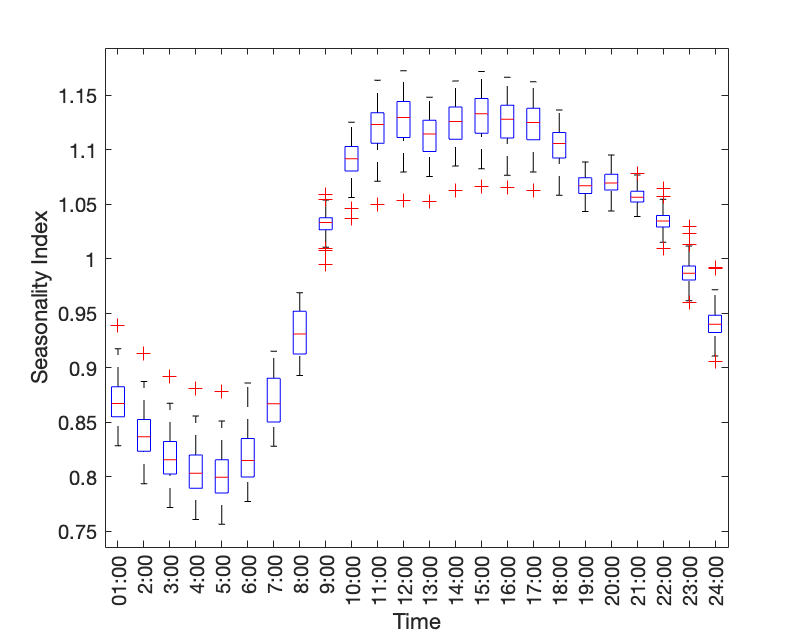}
			\caption{Monday}
			\label{Fig_Boxplots_SIMonday}
		\end{subfigure}
		\hfill
		\begin{subfigure}[b]{0.4\textwidth}
			\includegraphics[width=\textwidth]{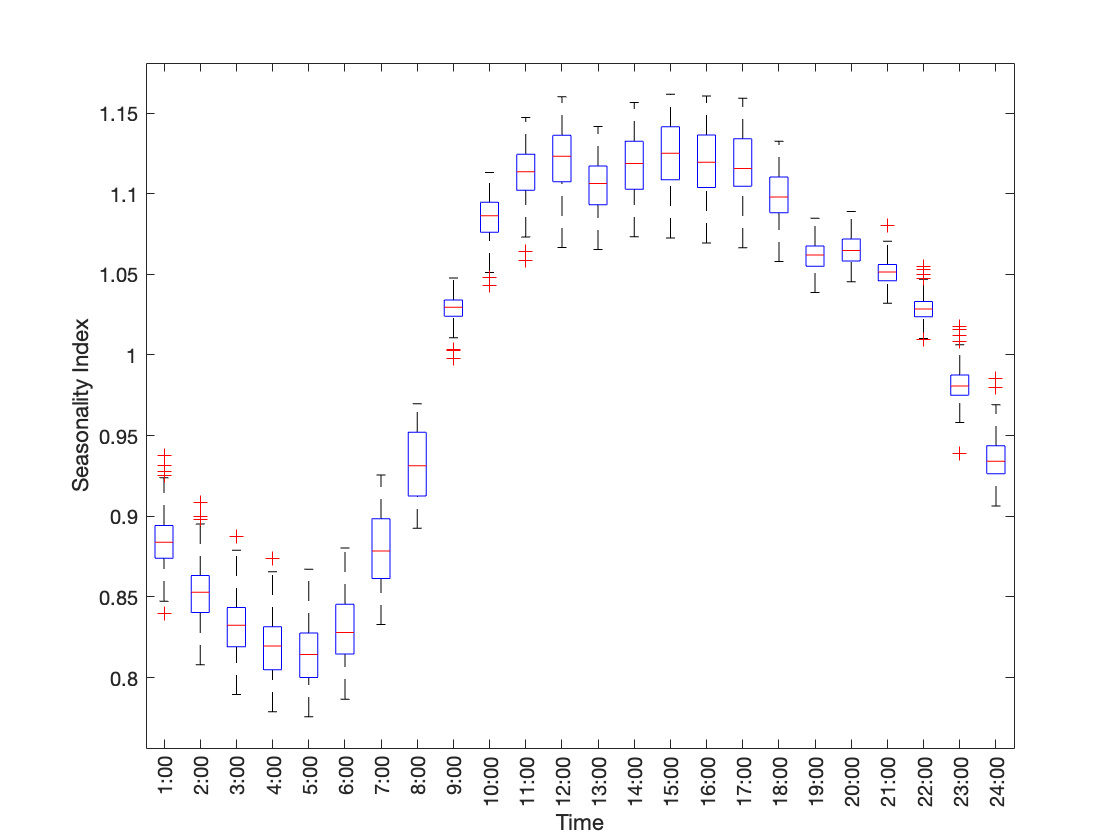}
			\caption{Tuesday}
			\label{Fig_Boxplots_SITuesday}
		\end{subfigure}
		\hfill
  \begin{subfigure}[b]{0.4\textwidth}
			\includegraphics[width=\textwidth]{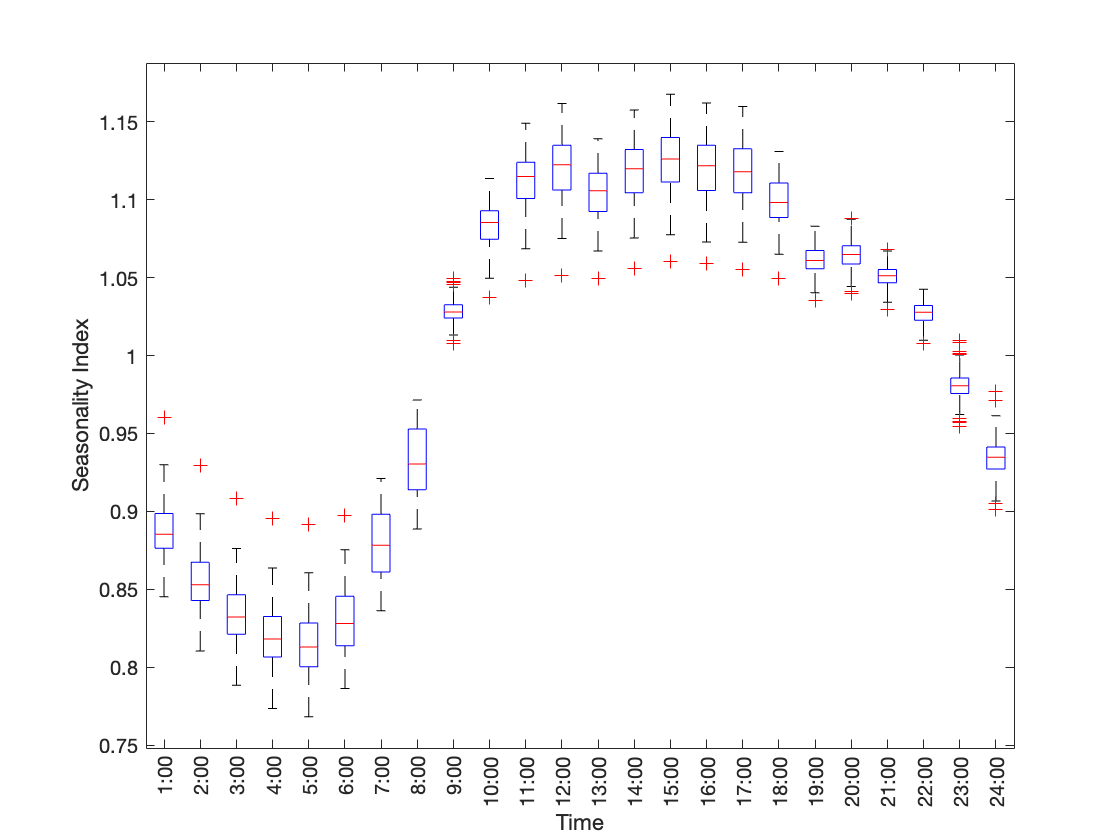}
			\caption{Wednesday}
			\label{Fig_Boxplots_SIWednesday}
		\end{subfigure}
		\hfill
  \begin{subfigure}[b]{0.4\textwidth}
			\includegraphics[width=\textwidth]{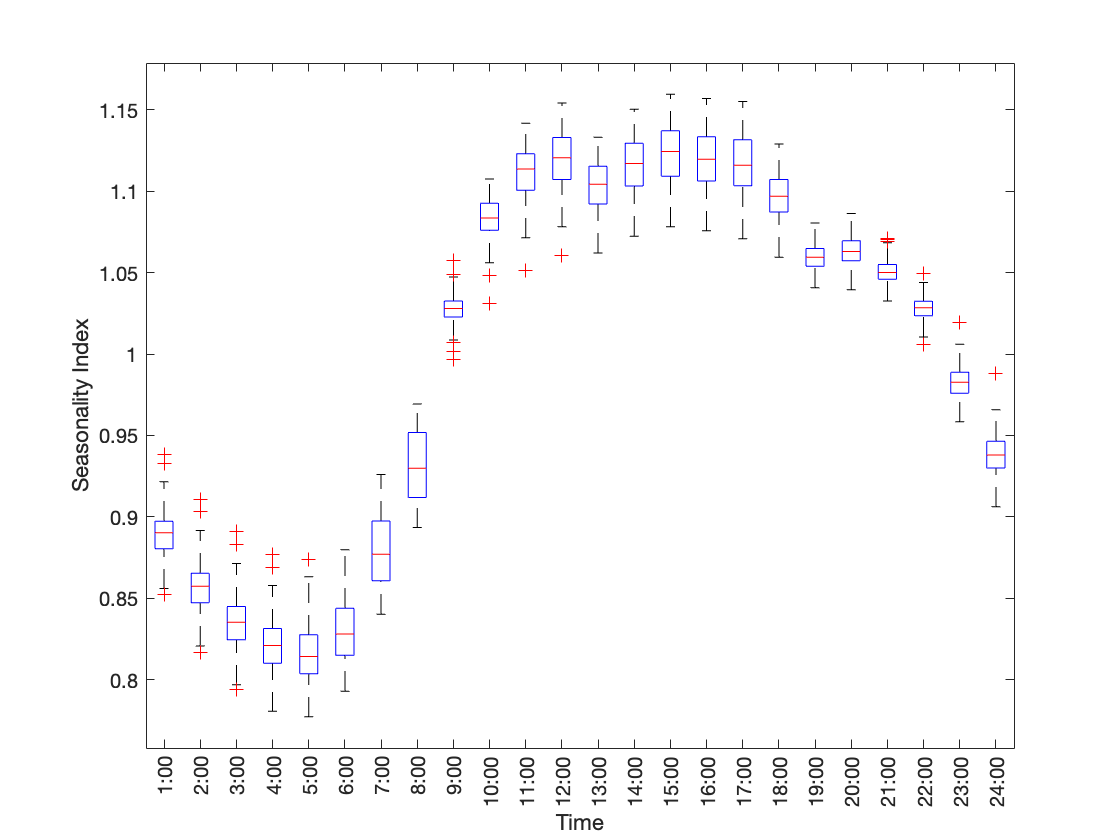}
			\caption{Thursday}
			\label{Fig_Boxplots_SIThursday}
		\end{subfigure}
		\hfill
  \begin{subfigure}[b]{0.4\textwidth}
			\includegraphics[width=\textwidth]{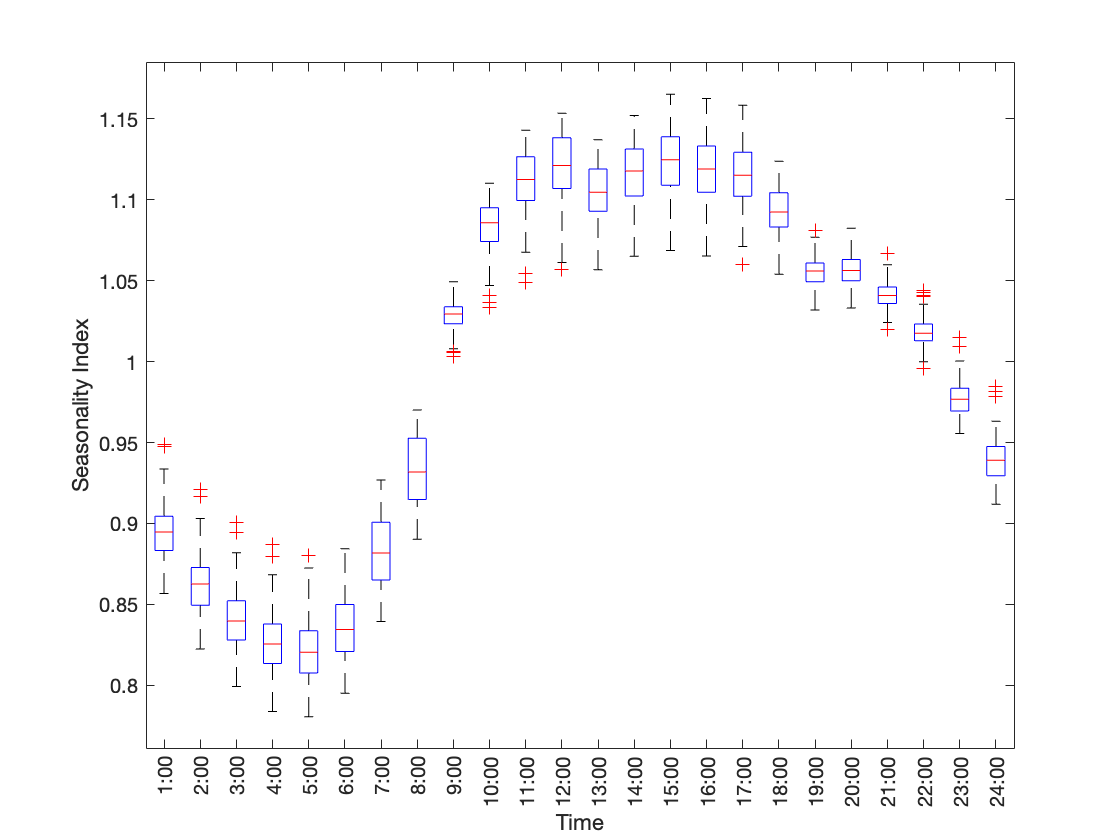}
			\caption{Friday}
			\label{Fig_Boxplots_SIFriday}
		\end{subfigure}
		\hfill
  \begin{subfigure}[b]{0.4\textwidth}
			\includegraphics[width=\textwidth]{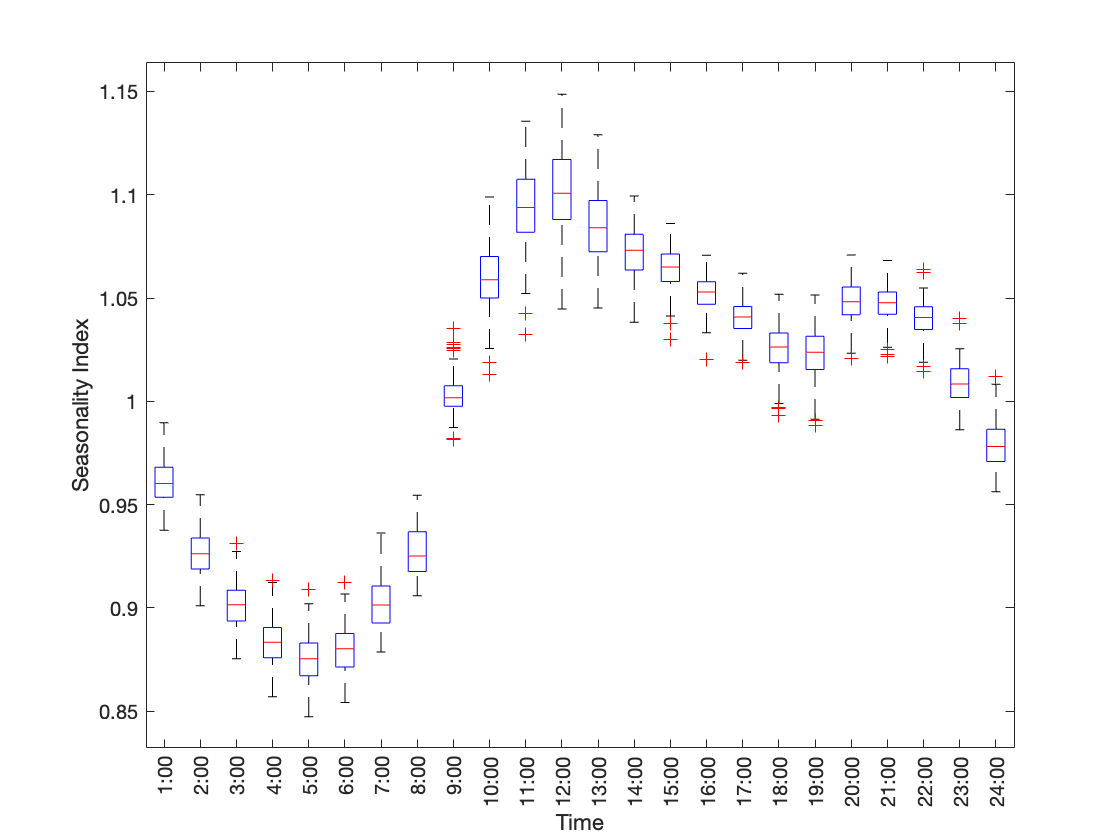}
			\caption{Saturday}
			\label{Fig_Boxplots_SISaturday}
		\end{subfigure}
		\hfill
  \begin{subfigure}[b]{0.4\textwidth}
			\includegraphics[width=\textwidth]{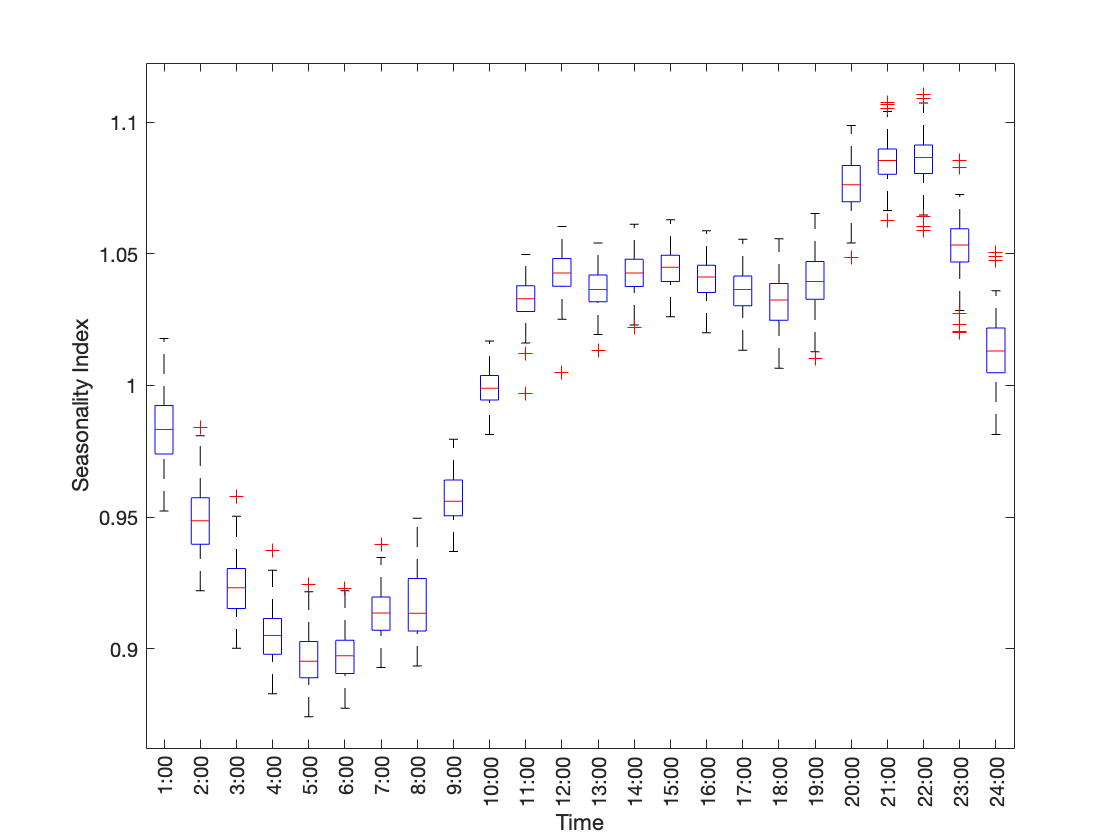}
			\caption{Sunday}
			\label{Fig_Boxplots_SISunday}
		\end{subfigure}
		\hfill
        \caption{\textbf{Box plots for Seasonality indices (2004-2022).} A box plot is shown for each hour of each day of the week.}
	\end{figure}

Figures \ref{Fig_Boxplots_SIMonday} to \ref{Fig_Boxplots_SISunday} show through box plots the range of SI values for every hour on each day of the week from 2004 to 2022. The small ranges again serve as observations to establishing stability in electricity demand. We can also see that during weekdays (Monday to Friday), the seasonality indices are similar to each other. This implies that demand exhibits the same behavior during weekdays.


\subsubsection{t-tests}\label{t-tests}

From the seasonality indices of each hour for each day of the week, month, and year combination, we can then check for stability of electricity demand. We proceed to show that for every hour of a specific day of the week, the difference in the seasonality indices across all possible pair-wise combinations remains small. To do so, we conduct a series of two-sample, one-tailed t-tests explained through an illustrative example. Suppose we are looking at the seasonality index of hour $h$ on day of the week $d$. Then, we take two samples: $SI_{y,m,d,h}^3$ -- the seasonality indices of hour $h$ on day of the week $d$ across all months $m$ of a year $y$, and $SI_{y',m,d,h}^3$ which is for year $y'$. Each sample would then have 12 data points, one for the average seasonality index for each month. Next, we let $\mu_y^{d, h} - \mu_{y'}^{d, h}$ be the population mean difference and let $\epsilon^{d, h} = 0$ be our hypothesis. We then conduct two two-sample, one tailed t-tests\footnote{It will be convenient in conducting the t-tests to let $\bar{x}_y^{d, h}$ and $s_y^{d, h}$ be the sample mean and standard deviation of the seasonality index of hour $h$, day $d$, and year $y$. That is, $\bar{x}_y^{d,h} = \frac{\sum_{m=1}^{12} SI_{y,m,d,h}^3}{12}$ and
    $\bar{s}_y^{d,h} = \frac{\sum_{m=1}^{12}(\bar{x}_y^{d,h} - SI_{y,m,d,h}^3)^2}{11}$}. The null hypothesis ($H_0^L$) for the left-tailed test is $\mu_y^{d, h} - \mu_{y'}^{d, h} \geq \epsilon^{d, h}$, with an alternative hypothesis ($H_1^L$) of $\mu_y^{d, h} - \mu_{y'}^{d, h} < \epsilon^{d,h}$. The null hypothesis ($H_0^R$) for the right-tailed test is $\mu_y - \mu_{y'} \leq -\epsilon^{d, h}$, with an alternative hypothesis ($H_1^R$) of $\mu_y^{d, h} - \mu_{y'}^{d, h} > \epsilon^{d, h}$. We increase $\epsilon^{d, h}$ by 0.001 every time one or both of the tests' null hypotheses are not rejected. Once both are rejected, then we stop. The final value of $\epsilon^{d, h}$ represents the one-sided maximum difference among the seasonality indices of hour $h$ on day of the week $d$ for the pair-wise years $y$ and $y'$. This is then repeated for every pairwise combination of years for hour $h$ and day of the week $d$, and the maximum across all pairs is taken as the final difference $\hat{\epsilon}^{d, h}$. Finally, the test is also repeated for every hour and day combination. Table \ref{Table_SIAbsoluteDifference} shows the resulting mean difference for each hour and day from 2004 to 2022.


\begin{table}[h]
\centering
\caption{Mean Difference of SIs for Each Hour and Day from 2004–2022 ($\hat{\epsilon}^{d, h}$)}
\label{Table_SIAbsoluteDifference}
\resizebox{\textwidth}{!}{%
\begin{tabular}{|c|c|c|c|c|c|c|c|}
\hline
\textbf{Time} & \textbf{Monday} & \textbf{Tuesday} & \textbf{Wednesday} & \textbf{Thursday} & \textbf{Friday} & \textbf{Saturday} & \textbf{Sunday} \\
\hline
\textbf{01:00} & 0.053 & 0.043 & 0.046 & 0.035 & 0.048 & 0.021 & 0.036 \\
\textbf{02:00} & 0.057 & 0.048 & 0.053 & 0.040 & 0.052 & 0.025 & 0.033 \\
\textbf{03:00} & 0.060 & 0.053 & 0.057 & 0.044 & 0.054 & 0.028 & 0.029 \\
\textbf{04:00} & 0.061 & 0.054 & 0.059 & 0.047 & 0.056 & 0.031 & 0.029 \\
\textbf{05:00} & 0.064 & 0.057 & 0.061 & 0.051 & 0.056 & 0.032 & 0.027 \\
\textbf{06:00} & 0.065 & 0.060 & 0.061 & 0.057 & 0.055 & 0.033 & 0.023 \\
\textbf{07:00} & 0.067 & 0.063 & 0.062 & 0.061 & 0.057 & 0.037 & 0.025 \\
\textbf{08:00} & 0.062 & 0.062 & 0.062 & 0.061 & 0.058 & 0.034 & 0.039 \\
\textbf{09:00} & 0.020 & 0.023 & 0.019 & 0.019 & 0.021 & 0.034 & 0.030 \\
\textbf{10:00} & 0.045 & 0.042 & 0.039 & 0.038 & 0.048 & 0.061 & 0.019 \\
\textbf{11:00} & 0.057 & 0.052 & 0.050 & 0.048 & 0.057 & 0.070 & 0.022 \\
\textbf{12:00} & 0.061 & 0.059 & 0.057 & 0.054 & 0.062 & 0.072 & 0.023 \\
\textbf{13:00} & 0.051 & 0.049 & 0.045 & 0.043 & 0.052 & 0.060 & 0.017 \\
\textbf{14:00} & 0.056 & 0.054 & 0.055 & 0.047 & 0.056 & 0.039 & 0.021 \\
\textbf{15:00} & 0.060 & 0.059 & 0.060 & 0.053 & 0.058 & 0.031 & 0.021 \\
\textbf{16:00} & 0.059 & 0.057 & 0.057 & 0.050 & 0.055 & 0.025 & 0.020 \\
\textbf{17:00} & 0.057 & 0.055 & 0.055 & 0.048 & 0.056 & 0.019 & 0.022 \\
\textbf{18:00} & 0.043 & 0.038 & 0.039 & 0.035 & 0.039 & 0.037 & 0.030 \\
\textbf{19:00} & 0.023 & 0.019 & 0.021 & 0.019 & 0.022 & 0.043 & 0.031 \\
\textbf{20:00} & 0.029 & 0.024 & 0.022 & 0.024 & 0.027 & 0.029 & 0.025 \\
\textbf{21:00} & 0.019 & 0.019 & 0.014 & 0.016 & 0.019 & 0.029 & 0.025 \\
\textbf{22:00} & 0.018 & 0.022 & 0.018 & 0.018 & 0.027 & 0.028 & 0.032 \\
\textbf{23:00} & 0.021 & 0.029 & 0.027 & 0.025 & 0.036 & 0.028 & 0.035 \\
\textbf{00:00} & 0.029 & 0.035 & 0.033 & 0.031 & 0.040 & 0.026 & 0.038 \\
\hline
\end{tabular}%
}
\end{table}

Next, we denote the mean difference as a percentage of the mean seasonality index for $h$ and $d$ as $\delta^{d, h}$. Mathematically, this is represented by:
\begin{equation}
    \delta^{d, h} = \frac{\hat{\epsilon}^{d, h}}{\mu^{d, h}}
\end{equation}
$\delta^{d, h}$ shows the largest percent deviation from the mean for a specific hour and day across all months and years. Table \ref{Table_SIPercentageDifference} shows the result for $\delta^{d, h}$ as well as the average for weekdays, weekends, and total week. The average $\delta^{d, h}$ across all hours and days is 4.24\%. During weekdays, this is slightly higher at 4.65\% lower at 3.21\% during weekends. What we can see is that aside from relatively larger deviations from the mean of 7-8\% during the 5:00 and 6:00 hours, $\delta^{d, h}$ remains small.

With both $\hat{\epsilon}^{d, h}$ and $\delta^{d, h}$, we have provided a measure through which we can say that Singapore's relative hour of the day seasonality indices for electricity load adjusted for monthly and day of the week seasonality has remained stable.

\begin{table}[h]
\centering
\caption{Mean Difference as a Percentage of Mean SI ($\delta^{d,h}$)}
\resizebox{\textwidth}{!}{%
\begin{tabular}{|c|c|c|c|c|c|c|c|c|c|c|}
\hline
\textbf{Time} & \textbf{Monday} & \textbf{Tuesday} & \textbf{Wednesday} & \textbf{Thursday} & \textbf{Friday} & \textbf{Saturday} & \textbf{Sunday} & \textbf{Weekday} & \textbf{Weekend} & \textbf{Week} \\
\hline
1:00  & 6.24 & 4.96 & 5.30 & 4.01 & 5.48 & 2.22 & 3.72 & 5.20 & 2.97 & 4.56 \\
2:00  & 6.96 & 5.74 & 6.34 & 4.76 & 6.16 & 2.75 & 3.54 & 5.99 & 3.14 & 5.18 \\
3:00  & 7.50 & 6.49 & 6.98 & 5.37 & 6.56 & 3.16 & 3.19 & 6.58 & 3.18 & 5.61 \\
4:00  & 7.73 & 6.72 & 7.34 & 5.82 & 6.91 & 3.57 & 3.25 & 6.90 & 3.41 & 5.91 \\
5:00  & 8.14 & 7.12 & 7.62 & 6.35 & 6.94 & 3.71 & 3.06 & 7.23 & 3.39 & 6.13 \\
6:00  & 8.06 & 7.32 & 7.45 & 6.95 & 6.68 & 3.80 & 2.60 & 7.29 & 3.20 & 6.12 \\
7:00  & 7.77 & 7.26 & 7.13 & 7.01 & 6.53 & 4.14 & 2.76 & 7.14 & 3.45 & 6.09 \\
8:00  & 6.71 & 6.72 & 6.70 & 6.59 & 6.26 & 3.69 & 4.28 & 6.60 & 3.99 & 5.85 \\
9:00  & 1.97 & 2.27 & 1.87 & 1.87 & 2.08 & 3.42 & 3.15 & 2.01 & 3.29 & 2.38 \\
10:00 & 4.22 & 3.94 & 3.65 & 3.56 & 4.51 & 5.84 & 1.92 & 3.97 & 3.88 & 3.95 \\
11:00 & 5.20 & 4.75 & 4.57 & 4.38 & 5.21 & 6.50 & 2.15 & 4.82 & 4.32 & 4.68 \\
12:00 & 5.52 & 5.35 & 5.17 & 4.89 & 5.62 & 6.63 & 2.23 & 5.31 & 4.43 & 5.06 \\
13:00 & 4.66 & 4.50 & 4.14 & 3.95 & 4.77 & 5.60 & 1.66 & 4.40 & 3.63 & 4.18 \\
14:00 & 5.08 & 4.91 & 5.01 & 4.27 & 5.09 & 3.68 & 2.04 & 4.87 & 2.86 & 4.30 \\
15:00 & 5.42 & 5.35 & 5.44 & 4.79 & 5.25 & 2.94 & 2.03 & 5.25 & 2.49 & 4.46 \\
16:00 & 5.34 & 5.18 & 5.18 & 4.53 & 5.00 & 2.40 & 1.94 & 5.05 & 2.17 & 4.23 \\
17:00 & 5.17 & 5.01 & 5.01 & 4.36 & 5.11 & 1.85 & 2.15 & 4.93 & 2.00 & 4.09 \\
18:00 & 3.96 & 3.51 & 3.61 & 3.23 & 3.62 & 3.64 & 2.94 & 3.59 & 3.29 & 3.50 \\
19:00 & 2.18 & 1.81 & 2.00 & 1.81 & 2.11 & 4.25 & 3.02 & 1.98 & 3.64 & 2.46 \\
20:00 & 2.74 & 2.28 & 2.09 & 2.28 & 2.59 & 2.80 & 2.35 & 2.40 & 2.58 & 2.45 \\
21:00 & 1.82 & 1.83 & 1.35 & 1.54 & 1.85 & 2.80 & 2.33 & 1.67 & 2.56 & 1.93 \\
22:00 & 1.76 & 2.16 & 1.77 & 1.77 & 2.68 & 2.72 & 2.98 & 2.03 & 2.85 & 2.26 \\
23:00 & 2.16 & 2.99 & 2.79 & 2.58 & 3.73 & 2.81 & 3.37 & 2.85 & 3.09 & 2.92 \\
00:00 & 3.14 & 3.80 & 3.59 & 3.36 & 4.32 & 2.69 & 3.81 & 3.64 & 3.25 & 3.53 \\
\hline
\textbf{Overall} & 4.98 & 4.67 & 4.67 & 4.17 & 4.79 & 3.65 & 2.77 & 4.65 & 3.21 & 4.24 \\
\hline
\end{tabular}%
}
\label{Table_SIPercentageDifference}
\end{table}

\section{Long-Term Load Forecasting Based on Stability Results}\label{Section_LTForecast}

We now demonstrate that the stability results for seasonality indices can be used to efficiently forecast long-term load for Singapore. Most of the methodologies for forecasting long-term hourly electricity load in Singapore are bottom-up approaches. We take a top-down approach and use GDP data as a statistically verifiable proxy for load growth. We then use the seasonality index results obtained in section \ref{Section_Seasonality_Index} to translate the total yearly forecast to an hourly forecast.

We provide 5-year ahead forecasts for hourly loads in 2017 to 2022. This assumes that each forecast year has available information up to the baseline years 2013 to 2018, respectively. This is important as 5-year ahead forecasts mean that information on actual GDP values are unavailable within the next five years. The accuracy of our hourly forecast also shows evidence of the predictive power of GDP for Singapore's electricity demand.

\subsection{Total Yearly Demand Forecast Using Linear Regression}
Simple linear regression models were shown to be effective in determining long-term electricity load \citeauthor{hammad_methods_2020} (\citeyear{hammad_methods_2020}).  Given this, we use simple linear regression to get an estimate of the total demand for a forecast year(dependent variable) from GDP(independent variable). GDP data starting 2004 is obtained from the yearly reports in the International Monetary Fund's World Economic Outlook Database (IMF WEO) (\citeauthor{IMF_2023}, \citeyear{IMF_2023}; the consolidated values are found in Table \ref{Table_IMF_GDP} in the Appendix.), while the total yearly demand is aggregated from the half-hourly data provided by the EMA (\citeauthor{ema_half-hourly_2023}, \citeyear{ema_half-hourly_2023}). Tables \ref{Table_LinearReg_2013to2015} and \ref{Table_LinearReg_2016to2018} show the linear regression models used to predict the total yearly load in baseline year $+$ 4.

\begin{table}[!h]
\caption{Linear Regression Models for Base Year 2013 to 2015}
\label{Table_LinearReg_2013to2015}
\centering
\resizebox{\textwidth}{!}{%
\begin{tabular}{|l|ccc|ccc|ccc|}
\hline
\textbf{Variable} & \multicolumn{3}{c|}{\textbf{2013}} & \multicolumn{3}{c|}{\textbf{2014}} & \multicolumn{3}{c|}{\textbf{2015}} \\
\cline{2-10}
 & \textbf{Coefficient} & \textbf{SE} & \textbf{p-value} & \textbf{Coefficient} & \textbf{SE} & \textbf{p-value} & \textbf{Coefficient} & \textbf{SE} & \textbf{p-value} \\
\hline
\textit{Intercept} & 60,170,382 & 1,450,000 & 0.00 & 61,109,978 & 1,270,000 & 0.00 & 60,914,095 & 1,260,000 & 0.00 \\
\textit{GDP (US \$)} & 123,370 & 7,298 & 0.00 & 115,679 & 5,936 & 0.00 & 116,795 & 5,598 & 0.00 \\
\hline
$R^2$ & 0.976 &  &  & 0.979 &  &  & 0.980 &  &  \\
\hline
\end{tabular}%
}
\end{table}

\begin{table}[!h]
\caption{Linear Regression Models for Base Year 2016 to 2018}
\label{Table_LinearReg_2016to2018}
\centering
\resizebox{\textwidth}{!}{%
\begin{tabular}{|l|ccc|ccc|ccc|}
\hline
\textbf{Variable} & \multicolumn{3}{c|}{\textbf{2016}} & \multicolumn{3}{c|}{\textbf{2017}} & \multicolumn{3}{c|}{\textbf{2018}} \\
\cline{2-10}
 & \textbf{Coefficient} & \textbf{SE} & \textbf{p-value} & \textbf{Coefficient} & \textbf{SE} & \textbf{p-value} & \textbf{Coefficient} & \textbf{SE} & \textbf{p-value} \\
\hline
\textit{Intercept} & 59,824,815 & 1,770,000 & 0.00 & 58,950,915 & 2,230,000 & 0.00 & 58,765,829 & 1,860,000 & 0.00 \\
\textit{GDP (US \$)} & 123,719 & 7,640 & 0.00 & 129,157 & 9,359 & 0.00 & 129,589 & 7,520 & 0.00 \\
\hline
$R^2$ & 0.963 &  &  & 0.945 &  &  & 0.961 &  &  \\
\hline
\end{tabular}%
}
\end{table}

The total load forecast is the sum of half-hourly load across the entire year. This is consistent with our data transformation of half-hourly load into its hourly equivalents. To get the total yearly consumption and the hourly consumption, we can divide the forecast values by two as explained in section \ref{Section_Data}. The $R^2$ values for the constructed linear regression models range from 0.945 to 0.980, indicating its high predictive capability.


\subsection{Conversion of Yearly Forecast to Hourly Forecast}

\begin{figure}[!h]
\centering
\includegraphics[width=\textwidth]{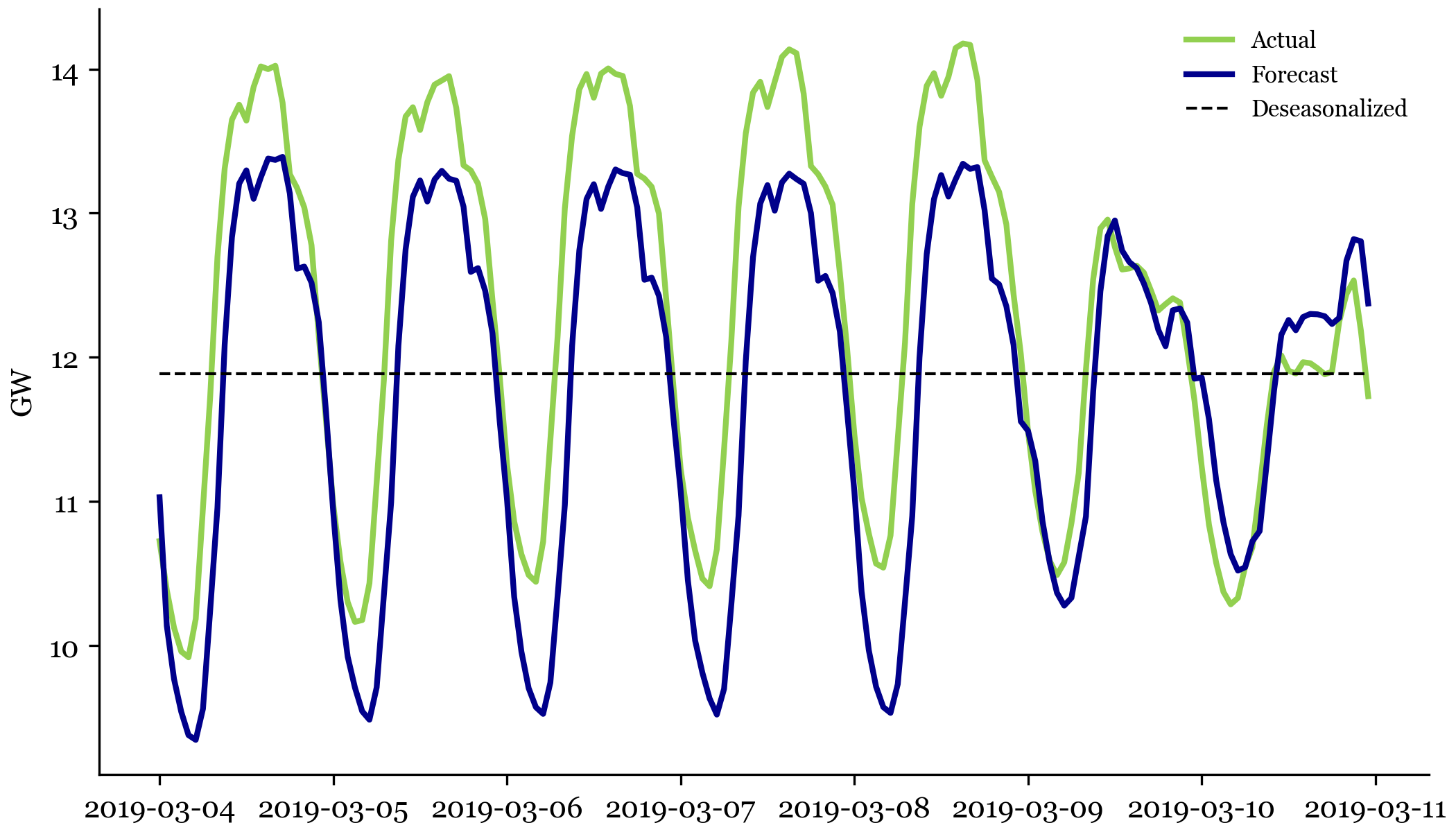}
            \caption{\textbf{Actual, Forecasted, and Deseasonalized Hourly Demand for March 4 to 10, 2019.} The baseline year is 2014.}
			\label{Fig_2019Forecast}
\end{figure}

The conversion from the yearly forecast to an hourly forecast makes use of the hourly seasonality indices from 5 years prior to the forecast year. That is, we divide the total yearly demand forecast by 365 days $\times$ 24 hours to get a de-seasonalized hourly demand. We then multiply the de-seasonalized hourly demand by the corresponding $SI_{y-4, m}^1$, $SI_{y-4, m,d}^2$, and $SI_{y-4, m, d, h}^3$ to get the final hourly forecast. Fig. \ref{Fig_2019Forecast} shows a week's sample of the de-seasonalized and final hourly forecast for 2019. The figure shows that even if we are looking at 5-year ahead forecasts, the forecasted hourly values still follow the general shape of the curve. This is because, as we have shown, the seasonality indices have remained stable such that indices from 5 years ago are still helpful in determining load curves\footnote{We do note that the peaks and throughs of the forecast during weekdays falls short of the actual values by a relatively constant amount each time. However, our forecast is still able to follow the general shape and trend of the actual load curves.}. Table \ref{Table_LongTermForecast_Results} shows the total yearly forecast, its corresponding percentage error, and the equivalent overall hourly percentage error (i.e. MAPE). The error of the total yearly demand forecast from 5-year prior data is less than 6.12\% across all years. Note however that this relatively larger deviation occured during recovery periods after the COVID-19 pandemic (2020 and 2021). If these are not accounted for, then the accuracy improves to 4.84\%. Similarly, the overall hourly errors are less than 6.87\% when including the years following COVID-19. If these are excluded, then the errors are less then 5.58\%. These values provide evidence for the effectiveness of our parsimonious, proposed methodology. 



While used a linear regression method here to obtain a yearly forecast, one can use any total yearly demand forecast and still utilize the stability methodology we developed to obtain a forecast of hourly resolution. For example, if we use the total yearly forecast of 2018 load provided by the EMA in it's 2017 SEMO of 106,000GW\footnote{The forecast is for 53000GWh, which we double to be consistent with our measures}, the hourly error for 2018 will be 4.9\% instead.

\begin{table}[h]
\centering
\caption{Singapore Long-Term Electricity Load Forecast (in MW)}
\label{Table_LongTermForecast_Results}
\begin{tabular}{|c|c|c|c|c|}
\hline
\textbf{Year} & \textbf{Actual} & \textbf{Forecast} & \textbf{Percentage Error (Total Year)} & \textbf{Percentage Error (Hourly)} \\
\hline
2017 & 103,147,207 & 101,485,969 & 1.61\% & 5.11\% \\
2018 & 107,482,983 & 102,284,277 & 4.84\% & 5.46\% \\
2019 & 107,965,323 & 104,135,499 & 3.55\% & 5.58\% \\
2020 & 105,666,163 & 102,751,430 & 2.76\% & 6.03\% \\
2021 & 111,352,369 & 104,536,553 & 6.12\% & 6.87\% \\
2022 & 113,587,819 & 112,539,730 & 0.92\% & 4.35\% \\
\hline
\end{tabular}
\end{table}


\section{Application to Other Countries}\label{Section_ExtensionOtherCountries}

We extend the use of the methodology to two other countries: Belgium and Bulgaria, chosen because of its status as an OECD and a non-OECD country, respectively. OECD countries exhibit well-established electricity infrastructure, with demand primarily driven by economic activity rather than by expanding access to electricity. As noted in \citeauthor{Steinbuks2017} (\citeyear{Steinbuks2017}), electricity consumption in OECD nations demonstrates a high but evolving correlation with GDP, where energy efficiency improvements and sectoral shifts toward services have gradually weakened the traditional relationship. Despite these changes, GDP remains a strong macroeconomic indicator of long-term electricity demand patterns. In non-OECD countries, electricity demand often tracks GDP more closely due to ongoing industrial growth and urbanization.  In these settings, economic expansion translates into increased electricity consumption, reflecting growing industrial output, rising household electrification rates, and infrastructure development. Applying our model to these economic contexts evaluates its viability in forecasting long-term electricity demand under different economic conditions.

Annual Historical GDP and electricity demand data for Belgium and Bulgaria were collected from \citeauthor{owid} (\citeyear{owid}) and \citeauthor{IMF_2023} (\citeyear{IMF_2023}). We calculated the seasonality indices and conducted the t-tests using the methdology established in \ref{t-tests} and found that load is also stable for both countries at 8.41\% and 13.24\% respectively. Pearson correlation coefficients were computed to establish the relationship between GDP and electricity demand, confirming a strong positive correlation in Belgium: 0.80 and Bulgaria: 0.92. Historical hourly electricity load data was obtained from \citeauthor{entsoe} (\citeyear{entsoe}) and the forecasting model was trained using the 2006–2017 dataset, employing a regression-based approach with GDP as the primary predictor variable. We then calculated the seasonality indices and generated two-year-ahead hourly forecasts for 2019, following the same long-term forecasting procedure as in the Singapore case study (Full details are in the Supplementary Material).

\begin{figure}[!h]
  \centering
\includegraphics[width=\textwidth]{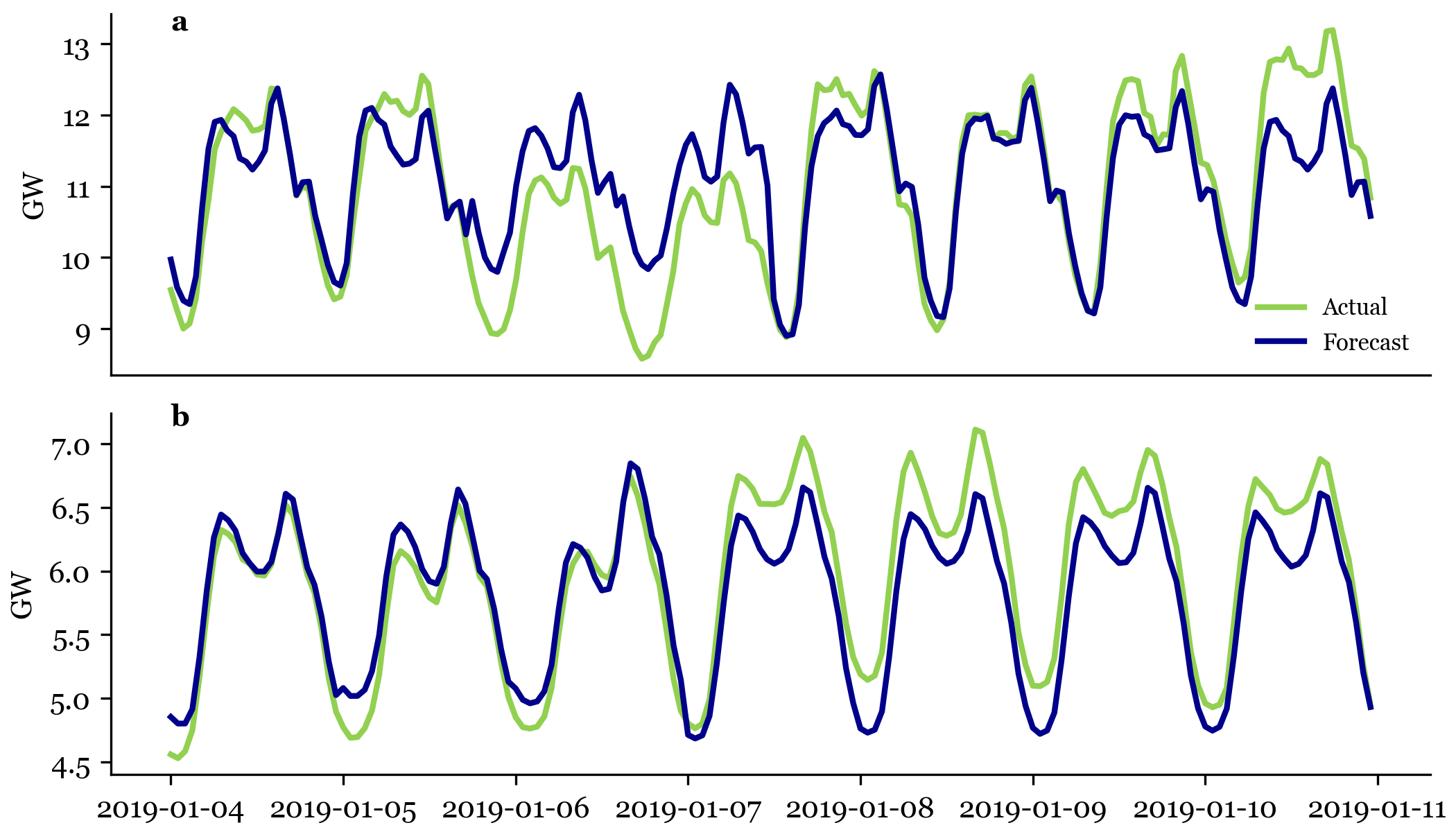}
            \caption{\textbf{Actual and Forecasted Hourly Demand for January 1 to 10, 2019 for Belgium (a) and Bulgaria (b).} The baseline year is 2017.}
            \label{Fig_BulgariaBelgium}
\end{figure}

The results are MAPEs of 6.81\% for Belgium and 5.64\% for Bulgaria. Fig. \ref{Fig_BulgariaBelgium}a and \ref{Fig_BulgariaBelgium}b shows the hourly forecast in a select week for Belgium and Bulgaria, respectively. Similar to the Singapore results, the forecasted load follows the shape of the actual load while enjoying low error values.

\section{Extension to Short-Term Load Forecasting}\label{Section_STForecast}

While we envision our methodology to be mainly used for long-term hourly load forecasting, we can nonetheless provide results when these are used for short-term forecasting. We use a triple seasonal model adopted from Holt-Winters' Exponential Smoothing model (\citeauthor{huang_electricity_2017}, \citeyear{huang_electricity_2017}). The notation for the Holt Winters' Exponential Smoothing model are level ($L_t)$, trend ($B_t$), seasons for hour of the day ($S_t^h$), day of the week ($S_t^d$), and month of the year ($S_t^m$) to get a Forecast $F_t$ for hour $t$. The initial trend $B_0$ is assumed to be 0 for the first year. The formulas for a Holt-Winters' Exponential Smoothing model are:

\begin{equation}
    L_t = L_{t-1} + B_{t-1} + \alpha\left(\frac{D_t}{S_{t-m_h}^{h}S_{t-m_d}^{d}S_{t-m_m}^{m}} - (L_{t-1} + B_{t-1}) \right)
\end{equation}

\begin{equation}
    S_t^i = S_{t-m_i}^i + \gamma^i\left(\frac{D_t}{L_t} - S_{t-m_i}^i\right) \text{ for } i = h,d, m
\end{equation}

\begin{equation}
    B_t = B_{t-1} + \beta(L_t - L_{t-1} - B_{t-1})
\end{equation}

\begin{equation}
    F_{t+1} = (L_t + B_t)(S_{t-m_h}^{h}S_{t-m_d}^{d}S_{t-m_m}^{m})
\end{equation}








In a typical exponential smoothing model, we would need to figure out what the smoothing constants $\alpha$, $\beta$, and $\gamma^i$ are. In our case, however, we have shown in section \ref{Section_Seasonality_Index} that the seasonality indices have not changed by much from 2004 to 2022. We then make use of the overall seasonality indices for month, day, and year $\mu^{d, h}$ from (\ref{Formula_OverallSeasonalityIndex}) as proxies for the seasonality index calculations in the ES model above. That is, we assume that the $S_t^i$ are constant. We therefore only need to search for the optimal $\alpha$ and $\beta$ values that give the lowest MAPE, which we determine as $\alpha = 0.19$ and $\beta = 0.88$ (at 0.01 increments).

\subsection{Comparison with Alternative Forecasting Methods}

We compare the Exponential Smoothing model with alternative forecasting models to evaluate the effectiveness of our approach. We use four machine learning models namely: FBProphet, SARIMAX, MSTL+autoARIMA, and Dynamic Harmonic Regression in a day-ahead forecast. The description of each machine learning model can be found in \ref{Appendix_ML}. From Table \ref{Table_MAPE}, we can see that the ES model performs at a comparable rate as the regression and machine learning models, where the ES consistently has lower error than the Machine Learning methods in all years except 2020.

\begin{table}[h]
\caption{Yearly MAPE Results per Forecasting Method (in \%)}
\centering
\label{Table_MAPE}
\begin{tabular}{|l|c@{}l|c@{}l|c@{}l|c@{}l|c@{}l|}
\hline
& \multicolumn{2}{c|}{\textbf{2018}} & \multicolumn{2}{c|}{\textbf{2019}} & \multicolumn{2}{c|}{\textbf{2020}} & \multicolumn{2}{c|}{\textbf{2021}} & \multicolumn{2}{c|}{\textbf{2022}} \\ \hline
ES (Constant SI) & 2.81 & & 3.32 & & 5.41 & & 3.52 & & 3.14 & \\ \hline
\multicolumn{1}{|c|}{\textbf{Machine Learning}} & \multicolumn{2}{l|}{} & \multicolumn{2}{l|}{} & \multicolumn{2}{l|}{} & \multicolumn{2}{l|}{} & \multicolumn{2}{l|}{} \\ \hline
FB Prophet & 3.65 & & 3.14 & & 4.35 & & 5.26 & & 6.15 & \\ \hline
SARIMAX & 6.22 & & 4.94 & & 4.41 & & 7.45 & & 5.40 & \\ \hline
MSTL + AutoARIMA & 3.35 & & 3.36 & & 4.49 & & 5.06 & & 2.85 & \\ \hline
Dynamic Harmonic Regression & 3.61 & & 3.56 & & 4.54 & & 6.42 & & 3.27 & \\ \hline
\end{tabular}
\end{table}

\section{Conclusion}\label{Section_Conclusion}

In this paper, we propose a new long-term, high-resolution electricity demand forecasting methodology that uses a top-down, statistically verifiable approach of load stability and GDP correlation with load. Using publicly available data from Singapore, we show through a series of t-tests that Singapore's hourly seasonality indices adjusted for monthly and day of the week seasonality have remained stable from 2004 to 2022.  We then make use of GDP forecasts from prior years and these seasonality indices to provide hourly 5-year ahead forecasts. Our results show that even with a 5-year ahead planning horizon, our stability results coupled with a parsimonious forecasting methodology provides an error rate of less than 6.87\% across the six forecast years. The resulting methodology allows electricity planners to create baseline load curves that can be used in more complex power systems and energy policy analysis. The stability methodology and the top-down forecasting approach allows for a simple and accurate way to establish long-term electricity demand for Singapore.  The approach is further validated by applying it to Belgium and Bulgaria. Overall, the proposed methodology provides a computationally efficient and interpretable forecasting framework. That is, the method only require historical hourly load data and GDP to create an accurate forecast, whereas other models need detailed assumptions on different electrification drivers. This gives our methodology an advantage over existing long-term hourly forecasting approaches.

We do point out some limitations of our methodology. While our results indicate that GDP-based forecasting generalizes well across different economies, certain structural and policy-driven factors may influence its effectiveness. In OECD countries, energy efficiency policies and electrification may contribute to electricity demand decoupling from GDP over time. Technological advancements, and a shift toward service-oriented sectors must be considered when applying GDP-driven models to long-term projections in developed economies. In non-OECD countries, rapid expansion of electricity access and industrial policy may enforce additional complexity. Rapid electrification initiatives, industrial sector transitions, and population growth may drive demand independently of GDP, necessitating supplementary economic variables to refine predictions. Nevertheless, the stability-driven GDP-based top-down aggregate forecasting approach used in this paper can still be used along with corrections that accounts for rapid electrification and efficient improvements, which should still be more effective than a bottom-up forecasting approach.  This is a topic for future research. A potential avenue for additional future research involves the development of hybrid forecasting models that integrate GDP alongside sectoral economic indicators, such as industrial output and electrification rates, to improve predictive accuracy.

\section*{Funding Source}
This research was funded by the National Research Foundation (NRF).

\section*{Acknowledgements}
This research is supported by the National Research Foundation, Prime Minister’s Office, Singapore under its Campus for Research Excellence and Technological Enterprise (CREATE) programme (NRF2022-ITC004-0001).

\bibliography{soumya.bib}
	
\newpage	
\appendix

\section{World Economic Outlook Database - GDP data for Singapore}
\begin{table}[h]
\centering
\caption{Actual (boldfaced) and forecasted GDP in report year Billion US \$}
\label{Table_IMF_GDP}
\begin{tabular}{|c|c|c|c|c|c|c|}
\hline
& \multicolumn{1}{c|}{\textbf{2013}} & \multicolumn{1}{c|}{\textbf{2014}} & \multicolumn{1}{c|}{\textbf{2015}} & \multicolumn{1}{c|}{\textbf{2016}} & \multicolumn{1}{c|}{\textbf{2017}} & \multicolumn{1}{c|}{\textbf{2018}} \\ \hline
2004 & \textbf{112.70} & \textbf{114.19} & \textbf{114.18} & \textbf{114.18} & \textbf{114.18} & \textbf{114.19} \\ \hline
2005 & \textbf{125.43} & \textbf{127.42} & \textbf{127.42} & \textbf{127.42} & \textbf{127.42} & \textbf{127.42} \\ \hline
2006 & \textbf{145.64} & \textbf{147.79} & \textbf{147.80} & \textbf{147.80} & \textbf{147.80} & \textbf{147.79} \\ \hline
2007 & \textbf{177.87} & \textbf{179.98} & \textbf{179.98} & \textbf{179.98} & \textbf{179.98} & \textbf{179.98} \\ \hline
2008 & \textbf{190.59} & \textbf{192.23} & \textbf{192.23} & \textbf{192.23} & \textbf{192.23} & \textbf{192.23} \\ \hline
2009 & \textbf{188.83} & \textbf{192.41} & \textbf{192.41} & \textbf{192.41} & \textbf{192.41} & \textbf{192.41} \\ \hline
2010 & \textbf{231.70} & \textbf{236.42} & \textbf{236.42} & \textbf{236.42} & \textbf{236.42} & \textbf{236.42} \\ \hline
2011 & \textbf{265.62} & \textbf{274.07} & \textbf{275.37} & \textbf{275.23} & \textbf{275.61} & \textbf{275.97} \\ \hline
2012 & \textbf{276.52} & \textbf{286.91} & \textbf{289.94} & \textbf{289.27} & \textbf{289.17} & \textbf{290.68} \\ \hline
2013 & 287.37 & \textbf{297.94} & \textbf{302.25} & \textbf{300.29} & \textbf{302.51} & \textbf{304.45} \\ \hline
2014 & 295.97 & 307.09 & \textbf{307.87} & \textbf{306.36} & \textbf{308.16} & \textbf{311.55} \\ \hline
2015 & 308.58 & 320.25 & 293.96 & \textbf{292.73} & \textbf{296.84} & \textbf{304.09} \\ \hline
2016 & 321.58 & 331.35 & 308.72 & 296.64 & \textbf{296.97} & \textbf{309.75} \\ \hline
2017 & 334.86 & 343.25 & 328.30 & 311.28 & 305.76 & \textbf{323.90} \\ \hline
2018 & 348.58 & 355.94 & 348.80 & 322.71 & 316.87 & 346.62 \\ \hline
2019 & & 369.07 & 370.06 & 334.71 & 328.42 & 359.62 \\ \hline
2020 & & & 394.98 & 347.07 & 340.41 & 377.23 \\ \hline
2021 & & & & 357.29 & 352.81 & 395.64 \\ \hline
2022 & & & & & 365.60 & 414.96 \\ \hline
\end{tabular}
\caption*{\scriptsize{Note: The bold faced numbers are actual GDP values, while those that are not are forecasts. The values are in US \$ of the baseline year.}}
\end{table}

\section{Our World in Data - Electricity Consumption and GDP of countries}
The data is publicly available at \citeauthor{owid} (\citeyear{owid}). 
\pagebreak

\subsection{Singapore}
\begin{table}[h]
\centering
\caption{Growth Of Electricity Consumption And GDP Over The Years in Singapore}
\label{Table_Electricity_GDP_Growth}
\renewcommand{\arraystretch}{1.2} 
\small 
\begin{tabular}{|c|c|c|c|c|}
\hline
\textbf{Year} & \makecell{\textbf{Electricity Demand} \\ (in TWh)} & \makecell{\textbf{GDP} \\ (in nominal USD)} & \makecell{\textbf{Normalised} \\ \textbf{Electricity Demand}} & \makecell{\textbf{Normalised} \\ \textbf{GDP}} \\ \hline
2000 & 31.67 & 1.52E+11 & 0.0000 & 0.0000E+00 \\ \hline
2001 & 33.06 & 1.53E+11 & 0.0550 & 1.28E-03 \\ \hline
2002 & 34.66 & 1.58E+11 & 0.1184 & 1.88E-02 \\ \hline
2003 & 35.32 & 1.60E+11 & 0.1446 & 2.63E-02 \\ \hline
2004 & 36.82 & 1.75E+11 & 0.2040 & 7.53E-02 \\ \hline
2005 & 38.22 & 1.91E+11 & 0.2594 & 1.24E-01 \\ \hline
2006 & 39.49 & 2.12E+11 & 0.3097 & 1.92E-01 \\ \hline
2007 & 41.13 & 2.38E+11 & 0.3747 & 2.77E-01 \\ \hline
2008 & 41.67 & 2.53E+11 & 0.3960 & 3.24E-01 \\ \hline
2009 & 41.80 & 2.56E+11 & 0.4012 & 3.36E-01 \\ \hline
2010 & 45.36 & 2.98E+11 & 0.5422 & 4.70E-01 \\ \hline
2011 & 46.00 & 3.20E+11 & 0.5675 & 5.41E-01 \\ \hline
2012 & 46.94 & 3.34E+11 & 0.6048 & 5.87E-01 \\ \hline
2013 & 47.97 & 3.50E+11 & 0.6455 & 6.39E-01 \\ \hline
2014 & 49.30 & 3.64E+11 & 0.6982 & 6.83E-01 \\ \hline
2015 & 50.28 & 3.75E+11 & 0.7370 & 7.18E-01 \\ \hline
2016 & 51.59 & 3.88E+11 & 0.7889 & 7.62E-01 \\ \hline
2017 & 52.28 & 4.06E+11 & 0.8162 & 8.19E-01 \\ \hline
2018 & 52.97 & 4.20E+11 & 0.8436 & 8.66E-01 \\ \hline
2019 & 54.14 & 4.26E+11 & 0.8899 & 8.84E-01 \\ \hline
2020 & 53.07 & 4.09E+11 & 0.8475 & 8.30E-01 \\ \hline
2021 & 55.79 & 4.46E+11 & 0.9552 & 9.48E-01 \\ \hline
2022 & 56.92 & 4.62E+11 & 1.0000 & 1.00E+00 \\ \hline
\end{tabular}
\end{table}

Pearson Correlation between Electrcity Demand and GDP is 0.9947
\pagebreak

\subsection{Belgium}

\begin{table}[h]
\centering
\caption{Growth Of Electricity Consumption And GDP Over The Years in Belgium}
\label{Table_Electricity_GDP_Growth_Belgium}
\renewcommand{\arraystretch}{1.2} 
\small 
\begin{tabular}{|c|c|c|c|c|}
\hline
\textbf{Year} & \makecell{\textbf{Electricity Demand} \\ (in TWh)} & \makecell{\textbf{GDP} \\ (in nominal USD)} & \makecell{\textbf{Normalised} \\ \textbf{Electricity Demand}} & \makecell{\textbf{Normalised} \\ \textbf{GDP}} \\ \hline
1990 & 66.57 & 2.73279E+11 & 0.0000 & 0.0000 \\ \hline
1991 & 69.34 & 2.78687E+11 & 0.1023 & 0.0250 \\ \hline
1992 & 71.55 & 2.83358E+11 & 0.1838 & 0.0466 \\ \hline
1993 & 72.31 & 2.81005E+11 & 0.2119 & 0.0357 \\ \hline
1994 & 75.32 & 2.90433E+11 & 0.3230 & 0.0793 \\ \hline
1995 & 77.59 & 2.97764E+11 & 0.4068 & 0.1131 \\ \hline
1996 & 79.33 & 3.02924E+11 & 0.4710 & 0.1370 \\ \hline
1997 & 81.15 & 3.14576E+11 & 0.5382 & 0.1908 \\ \hline
1998 & 83.48 & 3.2124E+11  & 0.6242 & 0.2216 \\ \hline
1999 & 84.22 & 3.33122E+11 & 0.6515 & 0.2765 \\ \hline
2000 & 87.12 & 3.45707E+11 & 0.7586 & 0.3347 \\ \hline
2001 & 87.74 & 3.49015E+11 & 0.7815 & 0.3500 \\ \hline
2002 & 88.53 & 3.55706E+11 & 0.8106 & 0.3809 \\ \hline
2003 & 89.98 & 3.58937E+11 & 0.8642 & 0.3958 \\ \hline
2004 & 90.69 & 3.72499E+11 & 0.8904 & 0.4585 \\ \hline
2005 & 90.68 & 3.8084E+11  & 0.8900 & 0.4970 \\ \hline
2006 & 93.50 & 3.90875E+11 & 0.9941 & 0.5434 \\ \hline
2007 & 93.59 & 4.04722E+11 & 0.9974 & 0.6074 \\ \hline
2008 & 93.66 & 4.08291E+11 & 1.0000 & 0.6239 \\ \hline
2009 & 87.92 & 3.9949E+11  & 0.7881 & 0.5832 \\ \hline
2010 & 93.56 & 4.11262E+11 & 0.9963 & 0.6376 \\ \hline
2011 & 91.77 & 4.20934E+11 & 0.9302 & 0.6823 \\ \hline
2012 & 91.45 & 4.24046E+11 & 0.9184 & 0.6967 \\ \hline
2013 & 91.59 & 4.25993E+11 & 0.9236 & 0.7057 \\ \hline
2014 & 88.80 & 4.32718E+11 & 0.8206 & 0.7368 \\ \hline
2015 & 89.43 & 4.41551E+11 & 0.8439 & 0.7776 \\ \hline
2016 & 90.50 & 4.47144E+11 & 0.8834 & 0.8034 \\ \hline
2017 & 91.36 & 4.54386E+11 & 0.9151 & 0.8369 \\ \hline
2018 & 91.21 & 4.62533E+11 & 0.9096 & 0.8746 \\ \hline
2019 & 90.84 & 4.72898E+11 & 0.8959 & 0.9225 \\ \hline
2020 & 87.95 & 4.47544E+11 & 0.7892 & 0.8053 \\ \hline
2021 & 91.51 & 4.74996E+11 & 0.9206 & 0.9322 \\ \hline
2022 & 86.61 & 4.89678E+11 & 0.7398 & 1.0000 \\ \hline
\end{tabular}
\end{table}

Pearson Correlation between Electrcity Demand and GDP is 0.8084
\pagebreak

\subsubsection{Graphs for Monthly and Hourly Demand for Belgium}

\begin{figure}[H]
		\caption{Monthly Demand for Belgium (in GW)}
  \centering
			\includegraphics[width=0.8\textwidth]{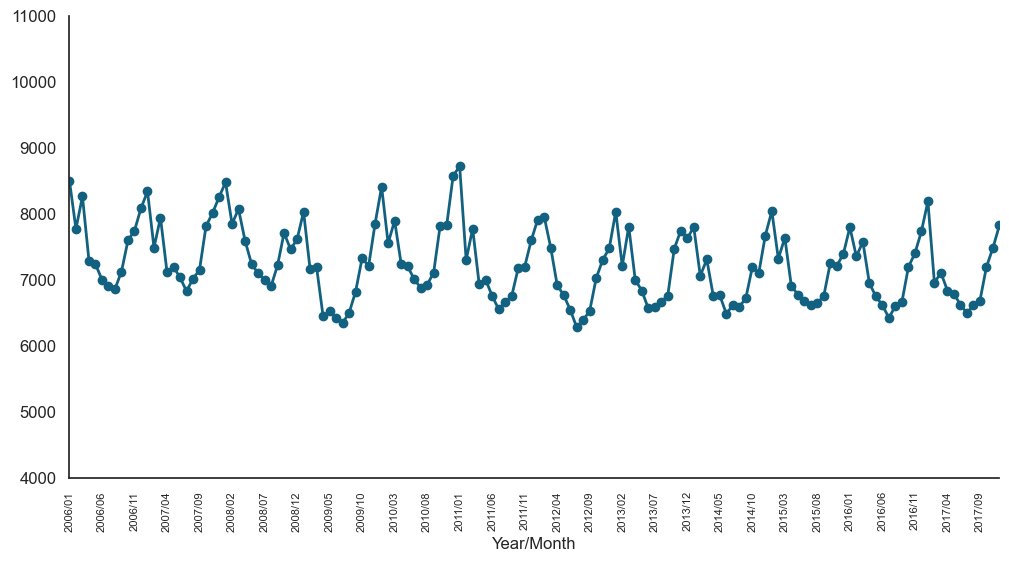}
			\label{Fig_MonthlyDemandBE}
		\hfill
\end{figure}

\vspace{100px}

\begin{figure}[H]
		\caption{Hourly Demand for Belgium (in MW)}
  \centering
			\includegraphics[width=0.8\textwidth]{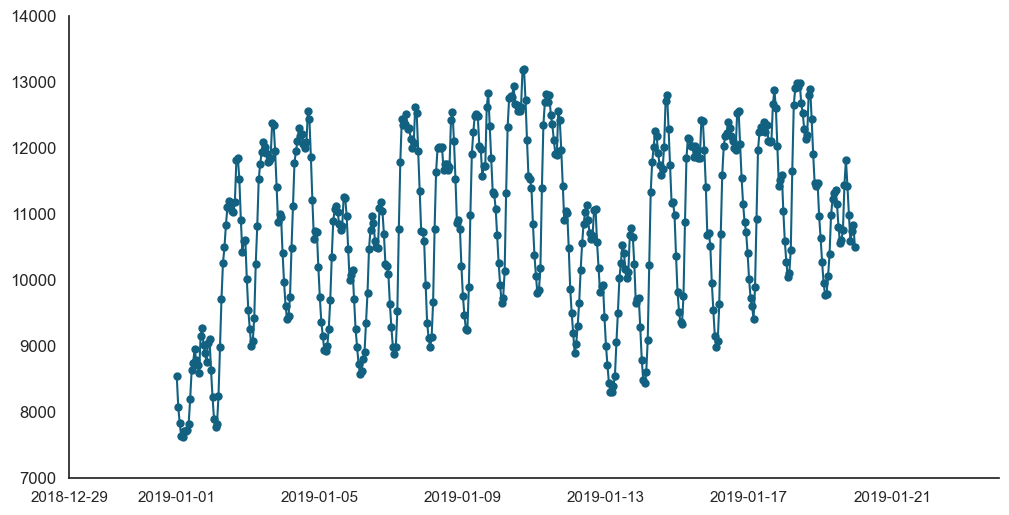}
			\label{Fig_HourlyDemand_SampleBE}
		\hfill
\end{figure}

\pagebreak

\subsubsection{t-Test Results for Belgium}

We replicate the t-tests for Belgium and find that load is also stable with an overall deviation across all days and hours of 8.41\% deviation.

\begin{table}[h]
\centering
\caption{Mean SI for each hour and day from 2006–2017 ($\mu^{d, h}$)}
\label{Table_MeanSI_BE}
\resizebox{\textwidth}{!}{%
\begin{tabular}{|c|c|c|c|c|c|c|c|}
\hline
\textbf{Time} & \multicolumn{1}{c|}{\textbf{Monday}} & \multicolumn{1}{c|}{\textbf{Tuesday}} & \multicolumn{1}{c|}{\textbf{Wednesday}} & \multicolumn{1}{c|}{\textbf{Thursday}} & \multicolumn{1}{c|}{\textbf{Friday}} & \multicolumn{1}{c|}{\textbf{Saturday}} & \multicolumn{1}{c|}{\textbf{Sunday}} \\ \hline
\textbf{0:00}  & 0.878 & 0.915 & 0.919 & 0.925 & 0.934 & 1.036 & 1.039 \\ \hline
\textbf{1:00}  & 0.836 & 0.868 & 0.874 & 0.879 & 0.889 & 0.978 & 0.982 \\ \hline
\textbf{2:00}  & 0.804 & 0.836 & 0.842 & 0.845 & 0.857 & 0.931 & 0.939 \\ \hline
\textbf{3:00}  & 0.791 & 0.821 & 0.829 & 0.830 & 0.842 & 0.907 & 0.907 \\ \hline
\textbf{4:00}  & 0.799 & 0.826 & 0.831 & 0.833 & 0.844 & 0.895 & 0.897 \\ \hline
\textbf{5:00}  & 0.838 & 0.858 & 0.861 & 0.861 & 0.874 & 0.889 & 0.892 \\ \hline
\textbf{6:00}  & 0.929 & 0.943 & 0.938 & 0.941 & 0.953 & 0.904 & 0.896 \\ \hline
\textbf{7:00}  & 1.005 & 1.016 & 1.013 & 1.013 & 1.027 & 0.925 & 0.899 \\ \hline
\textbf{8:00}  & 1.047 & 1.051 & 1.050 & 1.049 & 1.062 & 0.974 & 0.933 \\ \hline
\textbf{9:00}  & 1.079 & 1.076 & 1.072 & 1.071 & 1.083 & 1.027 & 0.985 \\ \hline
\textbf{10:00} & 1.094 & 1.086 & 1.084 & 1.082 & 1.093 & 1.057 & 1.023 \\ \hline
\textbf{11:00} & 1.116 & 1.105 & 1.104 & 1.102 & 1.110 & 1.078 & 1.060 \\ \hline
\textbf{12:00} & 1.097 & 1.082 & 1.084 & 1.078 & 1.087 & 1.069 & 1.069 \\ \hline
\textbf{13:00} & 1.094 & 1.079 & 1.081 & 1.076 & 1.079 & 1.044 & 1.034 \\ \hline
\textbf{14:00} & 1.083 & 1.072 & 1.071 & 1.069 & 1.065 & 1.022 & 1.003 \\ \hline
\textbf{15:00} & 1.076 & 1.065 & 1.066 & 1.064 & 1.050 & 1.010 & 0.987 \\ \hline
\textbf{16:00} & 1.072 & 1.061 & 1.057 & 1.059 & 1.040 & 1.008 & 0.981 \\ \hline
\textbf{17:00} & 1.089 & 1.077 & 1.071 & 1.073 & 1.056 & 1.033 & 1.015 \\ \hline
\textbf{18:00} & 1.090 & 1.073 & 1.068 & 1.072 & 1.053 & 1.051 & 1.054 \\ \hline
\textbf{19:00} & 1.071 & 1.052 & 1.049 & 1.051 & 1.028 & 1.043 & 1.070 \\ \hline
\textbf{20:00} & 1.042 & 1.023 & 1.022 & 1.022 & 0.999 & 1.017 & 1.065 \\ \hline
\textbf{21:00} & 1.015 & 1.000 & 0.998 & 0.996 & 0.974 & 1.003 & 1.063 \\ \hline
\textbf{22:00} & 1.041 & 1.021 & 1.020 & 1.019 & 1.006 & 1.049 & 1.113 \\ \hline
\textbf{23:00} & 1.015 & 0.995 & 0.994 & 0.994 & 0.994 & 1.049 & 1.094 \\ \hline
\end{tabular}}
\end{table}

\pagebreak

\begin{table}[h]
\centering
\caption{Mean difference of SIs for each hour and day from 2006–2017 ($\hat{\epsilon}^{d, h}$)}
\label{Table_SIAbsoluteDifference_BE}
\resizebox{\textwidth}{!}{%
\begin{tabular}{|c|c|c|c|c|c|c|c|}
\hline
\textbf{Time} & \multicolumn{1}{c|}{\textbf{Monday}} & \multicolumn{1}{c|}{\textbf{Tuesday}} & \multicolumn{1}{c|}{\textbf{Wednesday}} & \multicolumn{1}{c|}{\textbf{Thursday}} & \multicolumn{1}{c|}{\textbf{Friday}} & \multicolumn{1}{c|}{\textbf{Saturday}} & \multicolumn{1}{c|}{\textbf{Sunday}} \\ \hline
\textbf{00:00} & 0.103 & 0.099 & 0.099 & 0.099 & 0.095 & 0.132 & 0.143 \\ \hline
\textbf{01:00} & 0.072 & 0.068 & 0.070 & 0.068 & 0.066 & 0.100 & 0.109 \\ \hline
\textbf{02:00} & 0.057 & 0.052 & 0.050 & 0.049 & 0.053 & 0.067 & 0.080 \\ \hline
\textbf{03:00} & 0.086 & 0.074 & 0.068 & 0.069 & 0.071 & 0.053 & 0.058 \\ \hline
\textbf{04:00} & 0.153 & 0.142 & 0.136 & 0.136 & 0.138 & 0.057 & 0.056 \\ \hline
\textbf{05:00} & 0.198 & 0.192 & 0.186 & 0.190 & 0.191 & 0.094 & 0.073 \\ \hline
\textbf{06:00} & 0.153 & 0.154 & 0.150 & 0.152 & 0.155 & 0.124 & 0.099 \\ \hline
\textbf{07:00} & 0.094 & 0.088 & 0.090 & 0.089 & 0.089 & 0.139 & 0.123 \\ \hline
\textbf{08:00} & 0.063 & 0.062 & 0.059 & 0.058 & 0.060 & 0.114 & 0.119 \\ \hline
\textbf{09:00} & 0.047 & 0.051 & 0.050 & 0.049 & 0.047 & 0.080 & 0.101 \\ \hline
\textbf{10:00} & 0.048 & 0.052 & 0.042 & 0.047 & 0.050 & 0.051 & 0.075 \\ \hline
\textbf{11:00} & 0.088 & 0.090 & 0.077 & 0.086 & 0.085 & 0.067 & 0.073 \\ \hline
\textbf{12:00} & 0.075 & 0.068 & 0.063 & 0.072 & 0.076 & 0.082 & 0.105 \\ \hline
\textbf{13:00} & 0.071 & 0.067 & 0.062 & 0.071 & 0.075 & 0.068 & 0.087 \\ \hline
\textbf{14:00} & 0.059 & 0.057 & 0.048 & 0.057 & 0.058 & 0.046 & 0.056 \\ \hline
\textbf{15:00} & 0.039 & 0.038 & 0.034 & 0.033 & 0.037 & 0.046 & 0.054 \\ \hline
\textbf{16:00} & 0.049 & 0.045 & 0.041 & 0.042 & 0.040 & 0.080 & 0.108 \\ \hline
\textbf{17:00} & 0.049 & 0.051 & 0.047 & 0.048 & 0.051 & 0.072 & 0.116 \\ \hline
\textbf{18:00} & 0.082 & 0.078 & 0.076 & 0.074 & 0.076 & 0.058 & 0.081 \\ \hline
\textbf{19:00} & 0.088 & 0.078 & 0.079 & 0.077 & 0.073 & 0.064 & 0.062 \\ \hline
\textbf{20:00} & 0.065 & 0.056 & 0.057 & 0.054 & 0.044 & 0.045 & 0.054 \\ \hline
\textbf{21:00} & 0.047 & 0.045 & 0.043 & 0.040 & 0.025 & 0.034 & 0.053 \\ \hline
\textbf{22:00} & 0.121 & 0.117 & 0.115 & 0.111 & 0.103 & 0.103 & 0.136 \\ \hline
\textbf{23:00} & 0.143 & 0.137 & 0.138 & 0.132 & 0.139 & 0.151 & 0.163 \\ \hline
\end{tabular}}
\end{table}

\pagebreak

\begin{table}[h]
\centering
\caption{Mean difference as a percentage of Mean SI ($\delta^{d, h}$)}
\label{Table_SIPercentageDifference_BE}
\resizebox{\textwidth}{!}{%
\begin{tabular}{|c|c|c|c|c|c|c|c|c|c|c|}
\hline
\textbf{Time} & \multicolumn{1}{c|}{\textbf{Monday}} & \multicolumn{1}{c|}{\textbf{Tuesday}} & \multicolumn{1}{c|}{\textbf{Wednesday}} & \multicolumn{1}{c|}{\textbf{Thursday}} & \multicolumn{1}{c|}{\textbf{Friday}} & \multicolumn{1}{c|}{\textbf{Saturday}} & \multicolumn{1}{c|}{\textbf{Sunday}} & \multicolumn{1}{c|}{\textbf{Weekday}} & \multicolumn{1}{c|}{\textbf{Weekend}} & \multicolumn{1}{c|}{\textbf{Week}} \\ \hline
\textbf{00:00} & 11.73 & 10.83 & 10.77 & 10.70 & 10.17 & 12.74 & 13.76 & 10.84 & 13.25 & 11.53 \\ \hline
\textbf{01:00} & 8.62 & 7.83 & 8.01 & 7.73 & 7.42 & 10.22 & 11.10 & 7.92 & 10.66 & 8.70 \\ \hline
\textbf{02:00} & 7.09 & 6.22 & 5.94 & 5.80 & 6.19 & 7.19 & 8.52 & 6.25 & 7.86 & 6.71 \\ \hline
\textbf{03:00} & 10.88 & 9.01 & 8.21 & 8.32 & 8.43 & 5.84 & 6.39 & 8.97 & 6.12 & 8.15 \\ \hline
\textbf{04:00} & 19.16 & 17.20 & 16.36 & 16.33 & 16.35 & 6.37 & 6.25 & 17.08 & 6.31 & 14.00 \\ \hline
\textbf{05:00} & 23.63 & 22.38 & 21.61 & 22.07 & 21.84 & 10.57 & 8.19 & 22.31 & 9.38 & 18.61 \\ \hline
\textbf{06:00} & 16.46 & 16.34 & 15.98 & 16.16 & 16.26 & 13.72 & 11.05 & 16.24 & 12.38 & 15.14 \\ \hline
\textbf{07:00} & 9.35 & 8.66 & 8.89 & 8.79 & 8.67 & 15.02 & 13.68 & 8.87 & 14.35 & 10.44 \\ \hline
\textbf{08:00} & 6.02 & 5.90 & 5.62 & 5.53 & 5.65 & 11.71 & 12.75 & 5.74 & 12.23 & 7.59 \\ \hline
\textbf{09:00} & 4.36 & 4.74 & 4.66 & 4.58 & 4.34 & 7.79 & 10.25 & 4.53 & 9.02 & 5.82 \\ \hline
\textbf{10:00} & 4.39 & 4.79 & 3.88 & 4.35 & 4.58 & 4.82 & 7.33 & 4.39 & 6.08 & 4.88 \\ \hline
\textbf{11:00} & 7.88 & 8.15 & 6.98 & 7.81 & 7.66 & 6.22 & 6.89 & 7.69 & 6.55 & 7.37 \\ \hline
\textbf{12:00} & 6.84 & 6.29 & 5.81 & 6.68 & 6.99 & 7.67 & 9.83 & 6.52 & 8.75 & 7.16 \\ \hline
\textbf{13:00} & 6.49 & 6.21 & 5.74 & 6.60 & 6.95 & 6.51 & 8.41 & 6.40 & 7.46 & 6.70 \\ \hline
\textbf{14:00} & 5.45 & 5.32 & 4.48 & 5.33 & 5.45 & 4.50 & 5.59 & 5.21 & 5.04 & 5.16 \\ \hline
\textbf{15:00} & 3.62 & 3.57 & 3.19 & 3.10 & 3.52 & 4.55 & 5.47 & 3.40 & 5.01 & 3.86 \\ \hline
\textbf{16:00} & 4.57 & 4.24 & 3.88 & 3.97 & 3.84 & 7.94 & 11.01 & 4.10 & 9.47 & 5.64 \\ \hline
\textbf{17:00} & 4.50 & 4.74 & 4.39 & 4.47 & 4.83 & 6.97 & 11.43 & 4.59 & 9.20 & 5.90 \\ \hline
\textbf{18:00} & 7.52 & 7.27 & 7.12 & 6.90 & 7.22 & 5.52 & 7.69 & 7.21 & 6.60 & 7.03 \\ \hline
\textbf{19:00} & 8.22 & 7.42 & 7.53 & 7.33 & 7.10 & 6.14 & 5.79 & 7.52 & 5.97 & 7.07 \\ \hline
\textbf{20:00} & 6.24 & 5.47 & 5.58 & 5.28 & 4.41 & 4.42 & 5.07 & 5.40 & 4.75 & 5.21 \\ \hline
\textbf{21:00} & 4.63 & 4.50 & 4.31 & 4.02 & 2.57 & 3.39 & 4.99 & 4.00 & 4.19 & 4.06 \\ \hline
\textbf{22:00} & 11.63 & 11.46 & 11.27 & 10.90 & 10.24 & 9.82 & 12.22 & 11.10 & 11.02 & 11.07 \\ \hline
\textbf{23:00} & 14.09 & 13.76 & 13.88 & 13.28 & 13.98 & 14.40 & 14.90 & 13.80 & 14.65 & 14.04 \\ \hline
\midrule
\textbf{Overall} & 8.89 & 8.43 & 8.09 & 8.17 & 8.11 & 8.09 & 9.11 & 8.34 & 8.60 & 8.41 \\ \hline
\end{tabular}}
\end{table}

\pagebreak
\begin{figure}[H]
		\caption{Box plots for Seasonality indices (Belgium)}
		\begin{subfigure}[b]{0.4\textwidth}
			\includegraphics[width=\textwidth]{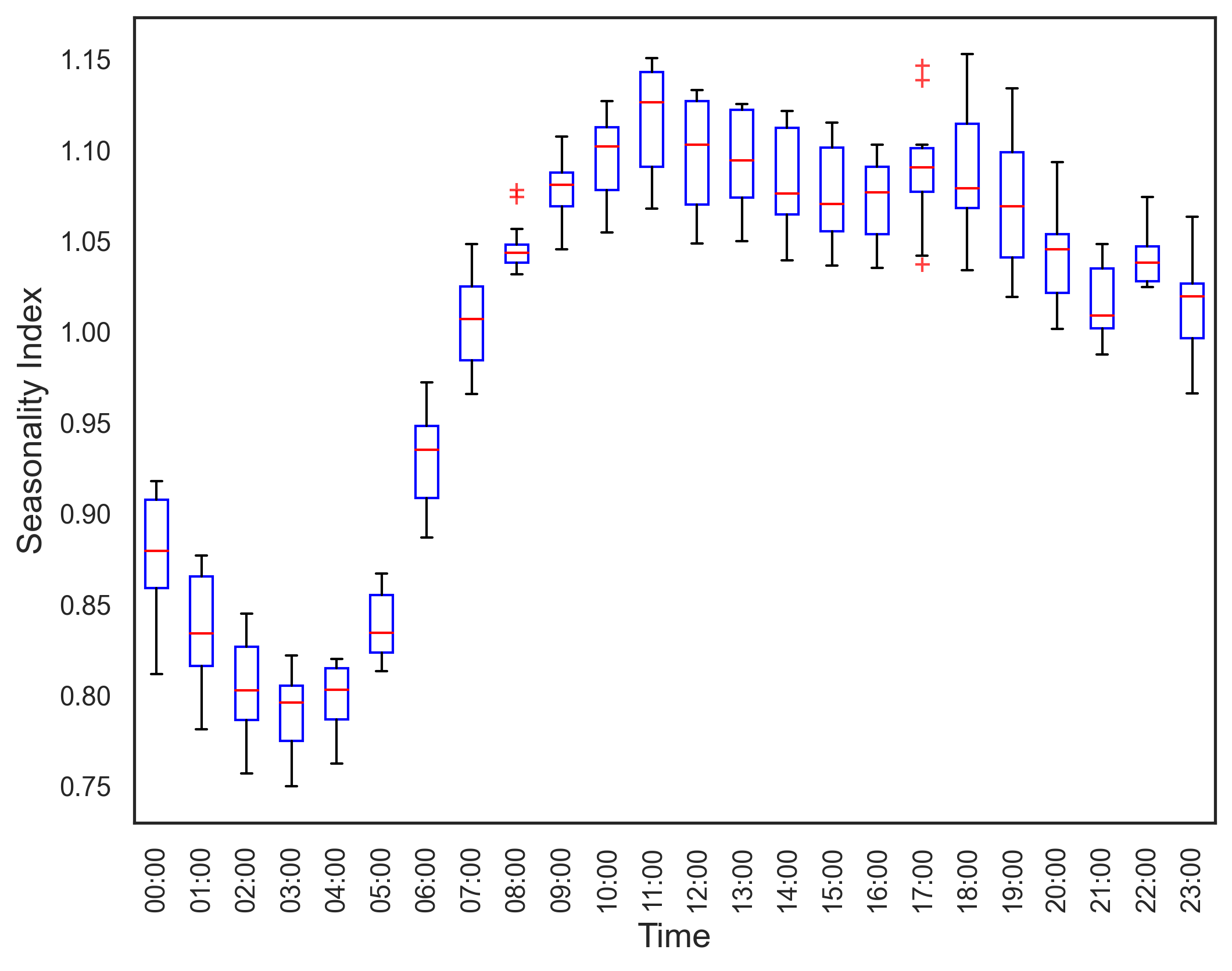}
			\caption{Monday}
			\label{Fig_Boxplots_SIMondayBE}
		\end{subfigure}
		\hfill
		\begin{subfigure}[b]{0.4\textwidth}
			\includegraphics[width=\textwidth]{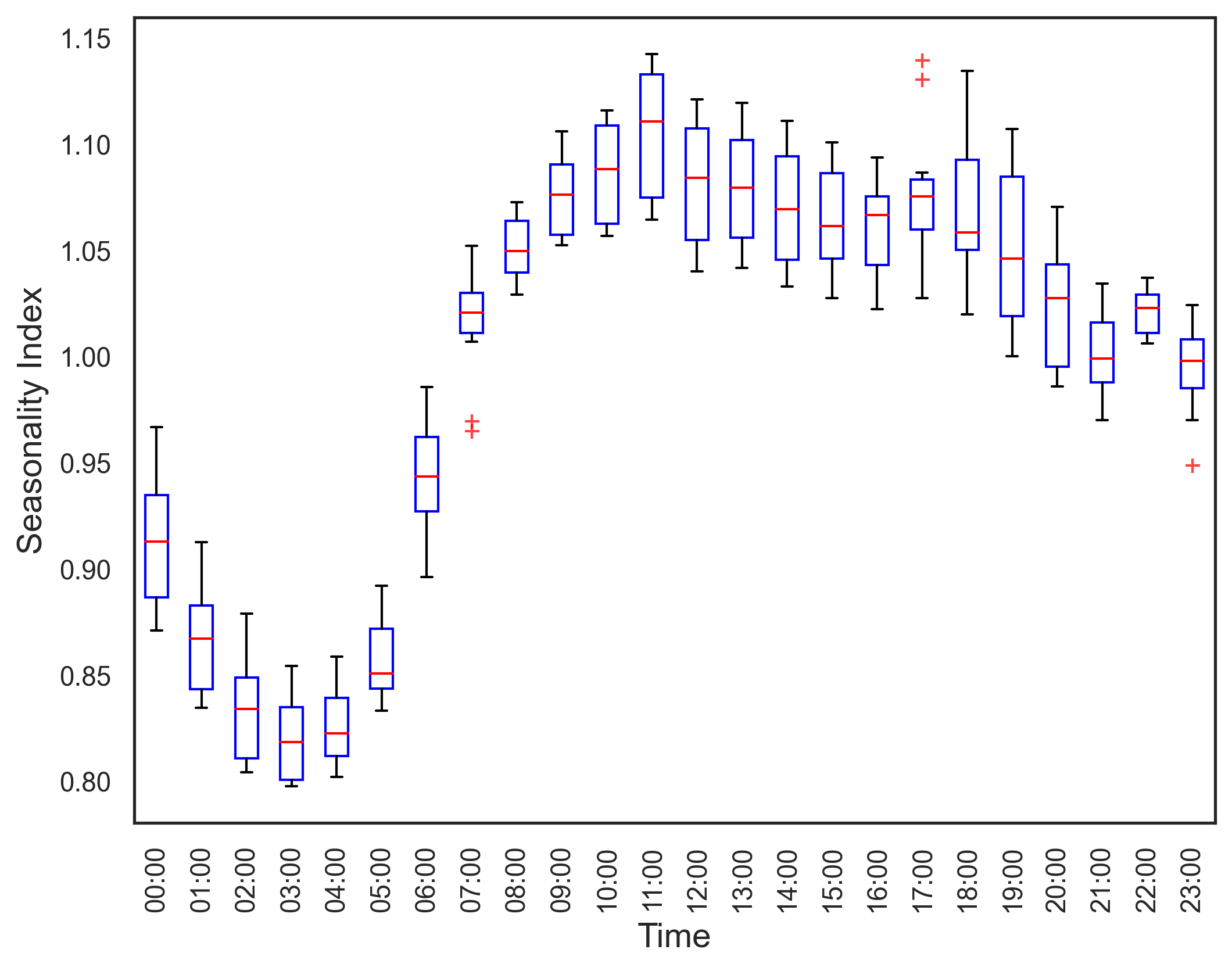}
			\caption{Tuesday}
			\label{Fig_Boxplots_SITuesdayBE}
		\end{subfigure}
		\hfill
  \begin{subfigure}[b]{0.4\textwidth}
			\includegraphics[width=\textwidth]{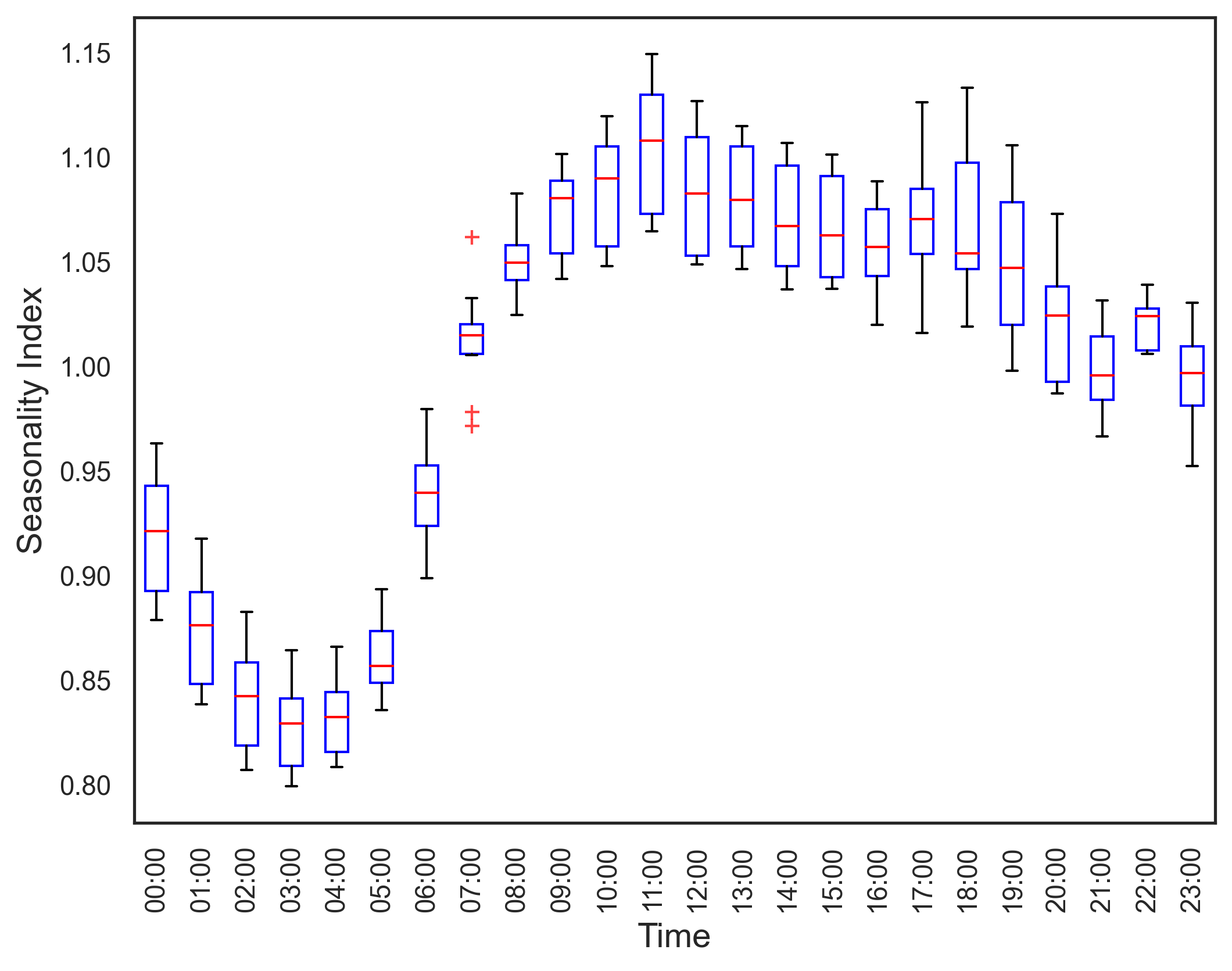}
			\caption{Wednesday}
			\label{Fig_Boxplots_SIWednesdayBE}
		\end{subfigure}
		\hfill
  \begin{subfigure}[b]{0.4\textwidth}
			\includegraphics[width=\textwidth]{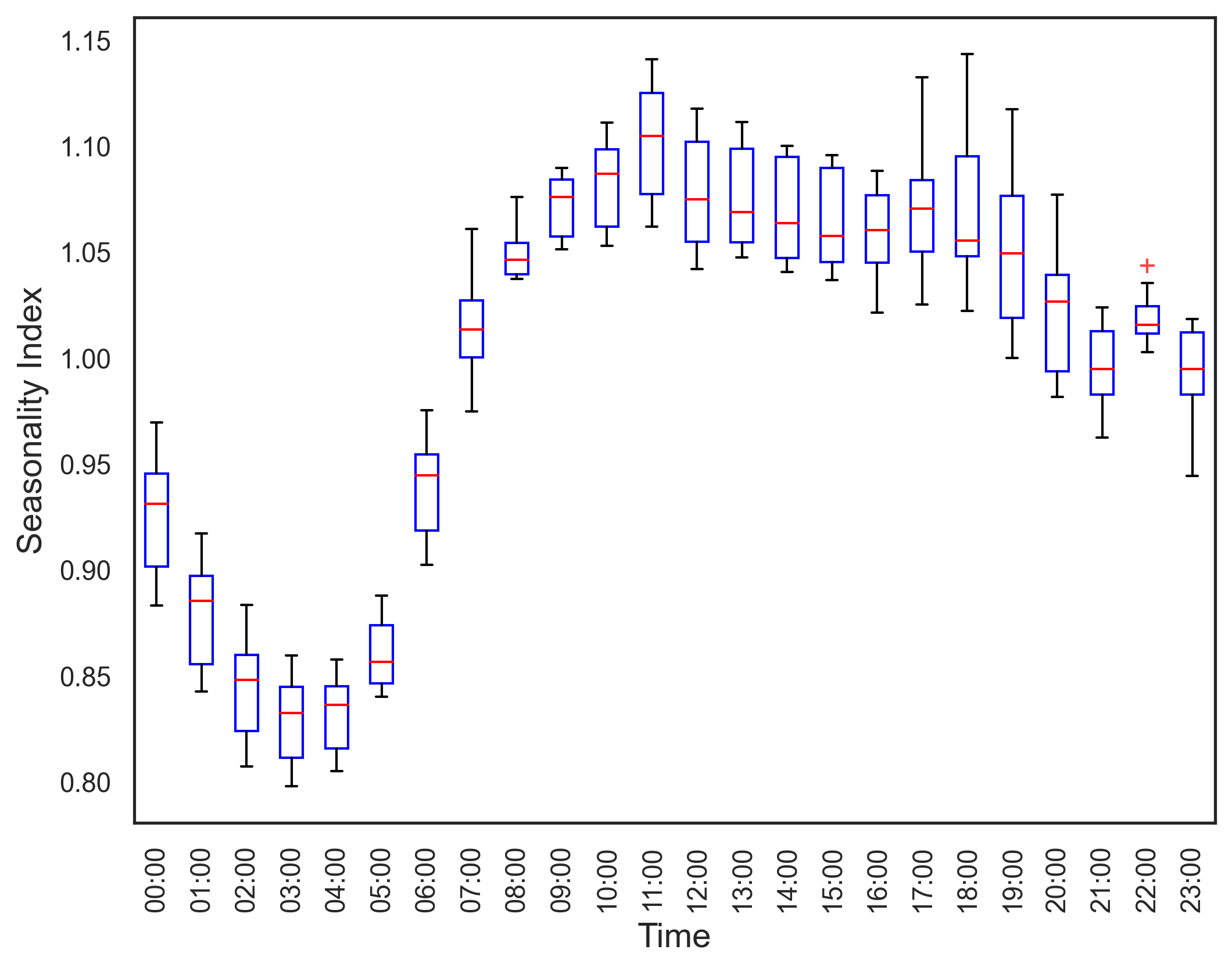}
			\caption{Thursday}
			\label{Fig_Boxplots_SIThursdayBE}
		\end{subfigure}
		\hfill
  \begin{subfigure}[b]{0.4\textwidth}
			\includegraphics[width=\textwidth]{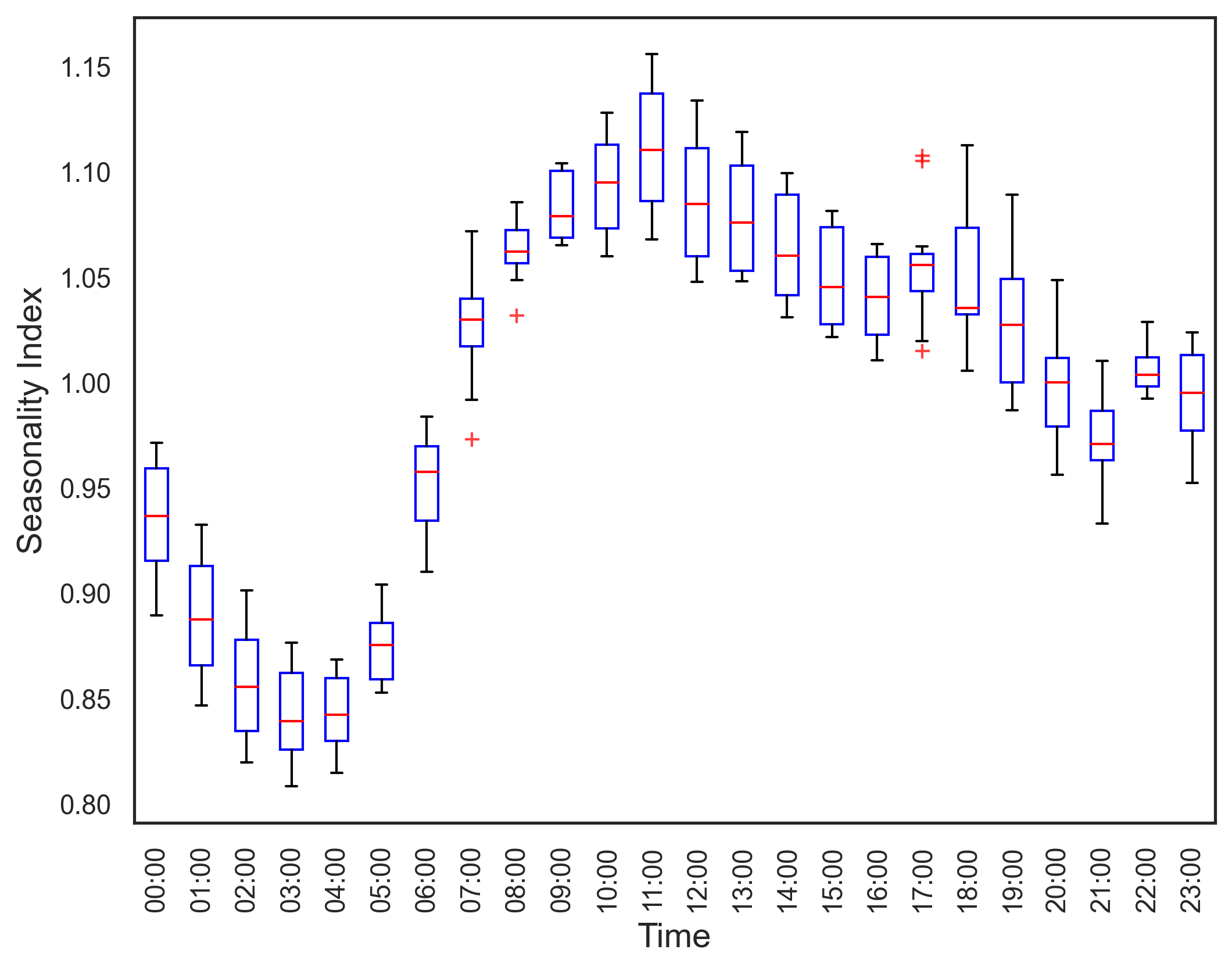}
			\caption{Friday}
			\label{Fig_Boxplots_SIFridayBE}
		\end{subfigure}
		\hfill
  \begin{subfigure}[b]{0.4\textwidth}
			\includegraphics[width=\textwidth]{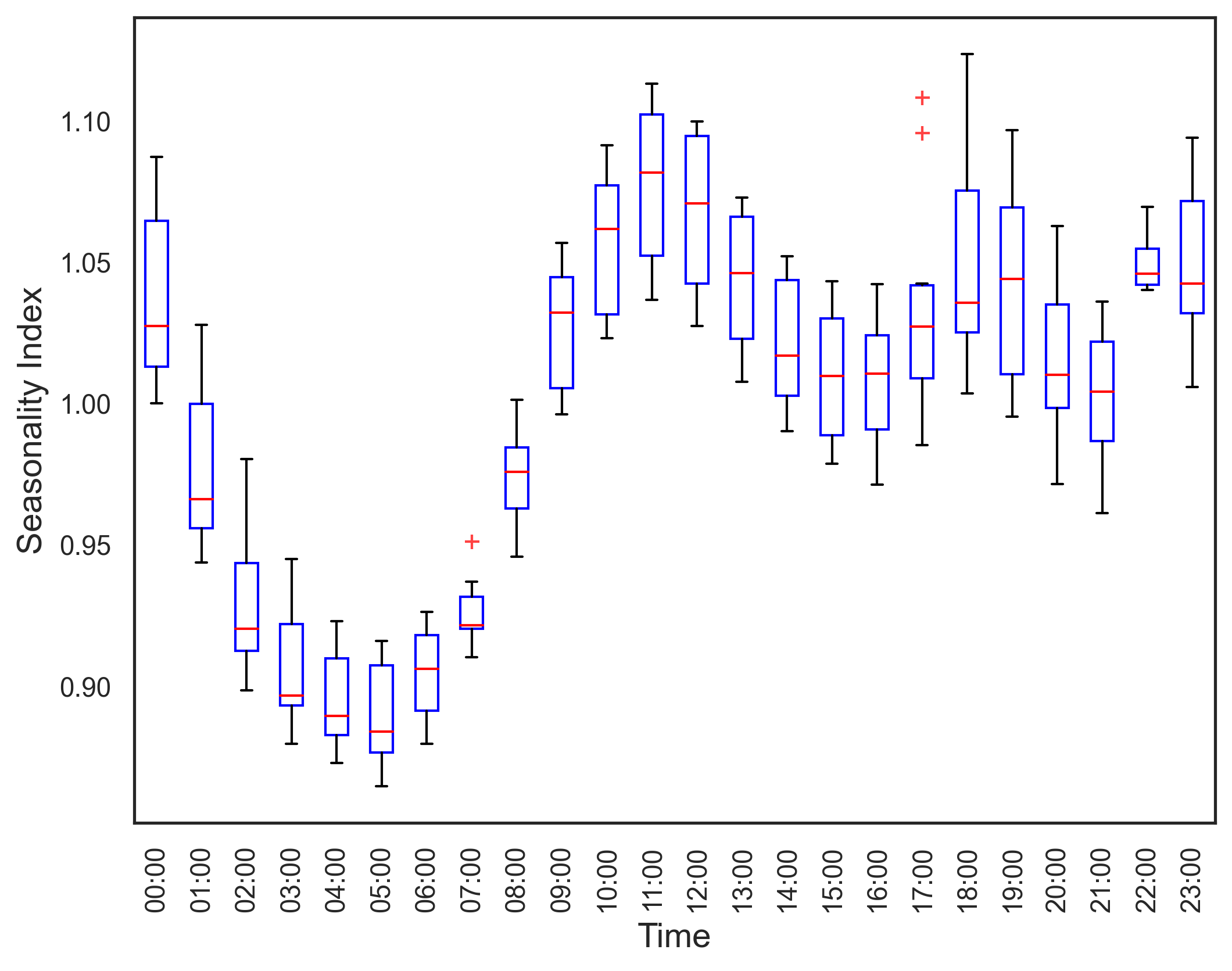}
			\caption{Saturday}
			\label{Fig_Boxplots_SISaturdayBE}
		\end{subfigure}
		\hfill
  \begin{subfigure}[b]{0.4\textwidth}
			\includegraphics[width=\textwidth]{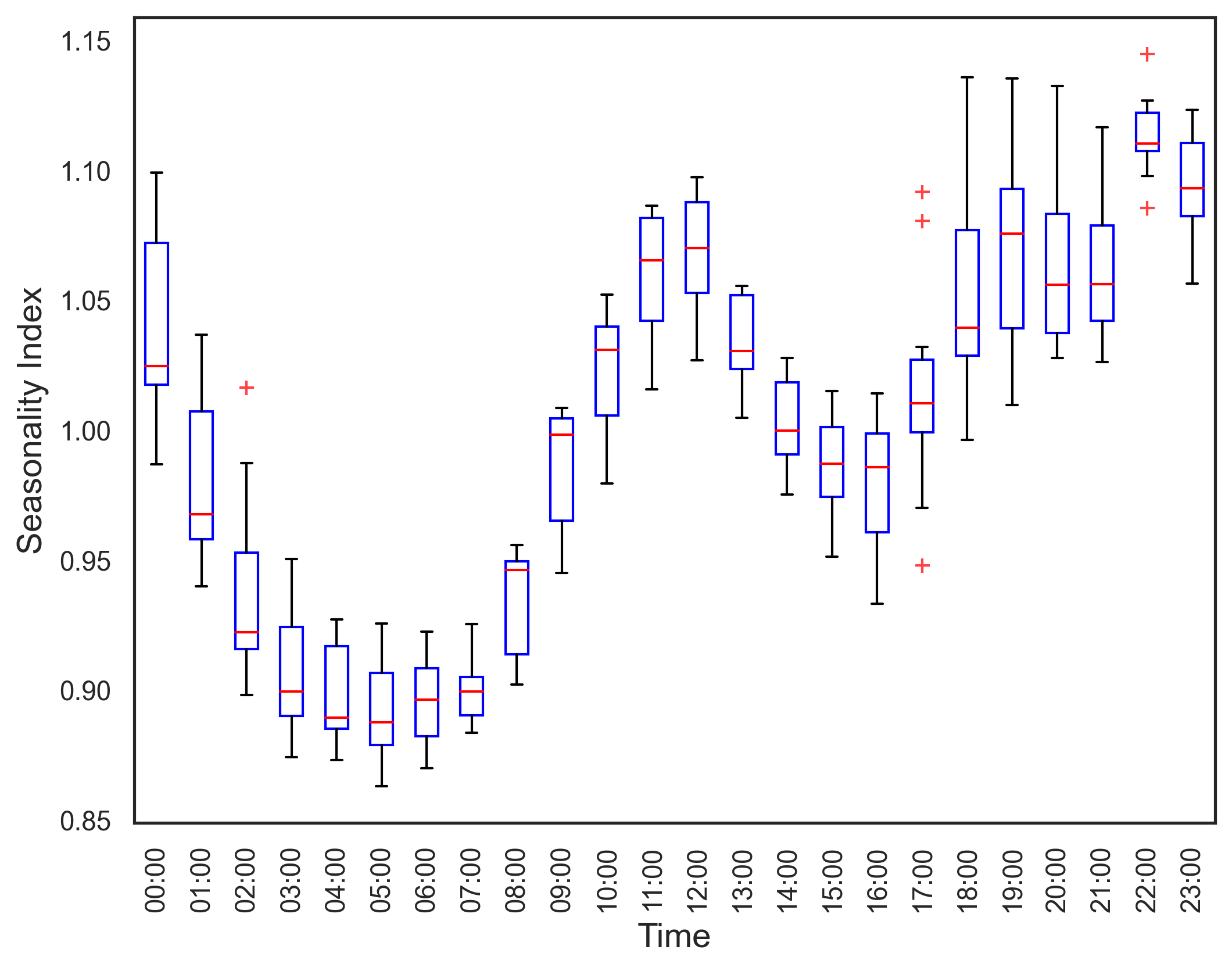}
			\caption{Sunday}
			\label{Fig_Boxplots_SISundayBE}
		\end{subfigure}
		\hfill
	\end{figure}

\pagebreak

\subsection{Bulgaria}

\begin{table}[h]
\centering
\caption{Growth Of Electricity Consumption And GDP Over The Years in Bulgaria}
\label{Table_Electricity_GDP_Growth_Bulgaria}
\renewcommand{\arraystretch}{1.2} 
\small 
\begin{tabular}{|c|c|c|c|c|}
\hline
\textbf{Year} & \makecell{\textbf{Electricity Demand} \\ (in TWh)} & \makecell{\textbf{GDP} \\ (in nominal USD)} & \makecell{\textbf{Normalised} \\ \textbf{Electricity Demand}} & \makecell{\textbf{Normalised} \\ \textbf{GDP}} \\ \hline
2000 & 33.31 & 9.14E+10 & 0.0375 & 0.0000 \\ \hline
2001 & 33.34 & 9.46E+10 & 0.0412 & 0.0344 \\ \hline
2002 & 33.01 & 9.83E+10 & 0.0000 & 0.0735 \\ \hline
2003 & 33.46 & 1.04E+11 & 0.0562 & 0.1345 \\ \hline
2004 & 34.45 & 1.14E+11 & 0.1798 & 0.2478 \\ \hline
2005 & 35.00 & 1.24E+11 & 0.2484 & 0.3484 \\ \hline
2006 & 36.17 & 1.35E+11 & 0.3945 & 0.4646 \\ \hline
2007 & 36.17 & 1.45E+11 & 0.3945 & 0.5714 \\ \hline
2008 & 36.89 & 1.58E+11 & 0.4844 & 0.7106 \\ \hline
2009 & 34.84 & 1.56E+11 & 0.2285 & 0.6938 \\ \hline
2010 & 37.60 & 1.66E+11 & 0.5730 & 0.8036 \\ \hline
2011 & 37.78 & 1.73E+11 & 0.5955 & 0.8775 \\ \hline
2012 & 38.41 & 1.76E+11 & 0.6742 & 0.9078 \\ \hline
2013 & 37.88 & 1.78E+11 & 0.6080 & 0.9260 \\ \hline
2014 & 38.06 & 1.81E+11 & 0.6305 & 0.9582 \\ \hline
2015 & 36.71 & 1.74E+11 & 0.4619 & 0.8839 \\ \hline
2016 & 36.59 & 1.69E+11 & 0.4469 & 0.8366 \\ \hline
2017 & 37.10 & 1.74E+11 & 0.5106 & 0.8828 \\ \hline
2018 & 38.00 & 1.79E+11 & 0.6230 & 0.9413 \\ \hline
2019 & 38.12 & 1.82E+11 & 0.6380 & 0.9691 \\ \hline
2020 & 38.04 & 1.80E+11 & 0.6280 & 0.9560 \\ \hline
2021 & 41.02 & 1.85E+11 & 1.0000 & 1.0000 \\ \hline
2022 & 38.23 & 1.76E+11 & 0.6517 & 0.9060 \\ \hline
\end{tabular}
\end{table}

Pearson Correlation between Electrcity Demand and GDP is 0.9204

\pagebreak

\subsubsection{Graphs for Monthly and Hourly Demand for Bulgaria}

\begin{figure}[H]
		\caption{Monthly Demand for Bulgaria (in MW)}
  \centering
			\includegraphics[width=0.8\textwidth]{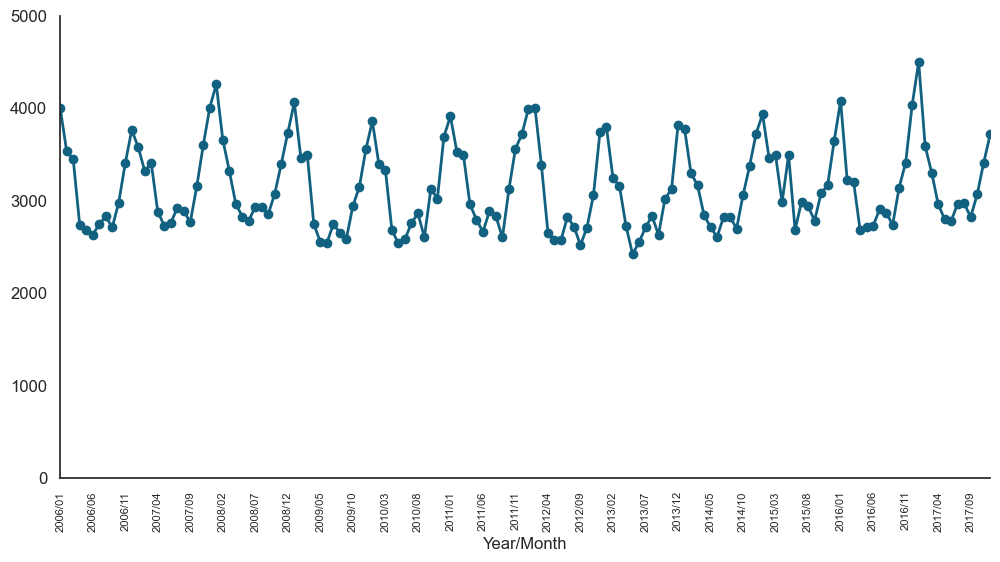}
			\label{Fig_MonthlyDemandBG}
		\hfill
\end{figure}

\vspace{100px}

\begin{figure}[H]
		\caption{Hourly Demand for Bulgaria (in KW)}
  \centering
			\includegraphics[width=0.8\textwidth]{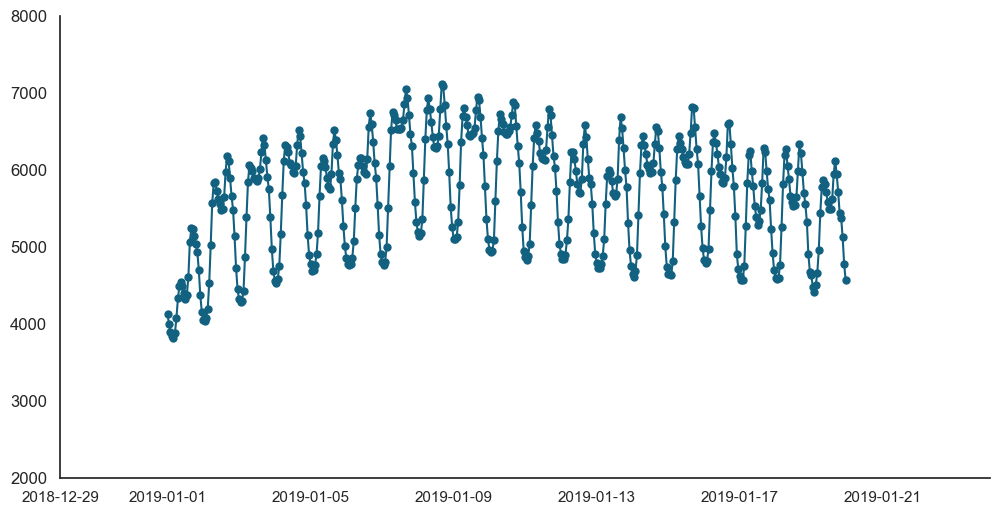}
			\label{Fig_HourlyDemand_SampleBG}
		\hfill
\end{figure}
\pagebreak
\subsubsection{t-Test Results for Bulgaria}

We replicate the t-tests for Bulgaria and find that load is also stable with an overall deviation across all days and hours of 13.24\% deviation.

\begin{table}[h]
\centering
\caption{Mean SI for each hour and day from 2006–2017 ($\mu^{d, h}$)}
\label{Table_MeanSI_BG}
\resizebox{\textwidth}{!}{%
\begin{tabular}{|c|c|c|c|c|c|c|c|}
\hline
\textbf{Time} & \multicolumn{1}{c|}{\textbf{Monday}} & \multicolumn{1}{c|}{\textbf{Tuesday}} & \multicolumn{1}{c|}{\textbf{Wednesday}} & \multicolumn{1}{c|}{\textbf{Thursday}} & \multicolumn{1}{c|}{\textbf{Friday}} & \multicolumn{1}{c|}{\textbf{Saturday}} & \multicolumn{1}{c|}{\textbf{Sunday}} \\ \hline
\textbf{0:00}  & 0.929 & 0.943 & 0.950 & 0.947 & 0.958 & 1.015 & 0.997 \\ \hline
\textbf{1:00}  & 0.858 & 0.875 & 0.881 & 0.879 & 0.887 & 0.939 & 0.926 \\ \hline
\textbf{2:00}  & 0.824 & 0.842 & 0.848 & 0.845 & 0.854 & 0.900 & 0.885 \\ \hline
\textbf{3:00}  & 0.808 & 0.826 & 0.831 & 0.832 & 0.837 & 0.878 & 0.859 \\ \hline
\textbf{4:00}  & 0.804 & 0.823 & 0.825 & 0.825 & 0.833 & 0.867 & 0.847 \\ \hline
\textbf{5:00}  & 0.832 & 0.844 & 0.848 & 0.847 & 0.852 & 0.872 & 0.851 \\ \hline
\textbf{6:00}  & 0.903 & 0.909 & 0.915 & 0.911 & 0.918 & 0.893 & 0.861 \\ \hline
\textbf{7:00}  & 0.980 & 0.978 & 0.981 & 0.983 & 0.984 & 0.933 & 0.885 \\ \hline
\textbf{8:00}  & 1.028 & 1.027 & 1.027 & 1.029 & 1.030 & 0.981 & 0.934 \\ \hline
\textbf{9:00}  & 1.060 & 1.051 & 1.049 & 1.052 & 1.054 & 1.019 & 0.986 \\ \hline
\textbf{10:00} & 1.058 & 1.048 & 1.044 & 1.047 & 1.048 & 1.029 & 1.013 \\ \hline
\textbf{11:00} & 1.058 & 1.046 & 1.041 & 1.043 & 1.049 & 1.035 & 1.030 \\ \hline
\textbf{12:00} & 1.055 & 1.041 & 1.033 & 1.037 & 1.043 & 1.034 & 1.035 \\ \hline
\textbf{13:00} & 1.057 & 1.044 & 1.039 & 1.044 & 1.043 & 1.020 & 1.019 \\ \hline
\textbf{14:00} & 1.045 & 1.033 & 1.027 & 1.032 & 1.028 & 0.996 & 0.994 \\ \hline
\textbf{15:00} & 1.041 & 1.034 & 1.026 & 1.029 & 1.024 & 0.986 & 0.993 \\ \hline
\textbf{16:00} & 1.038 & 1.031 & 1.026 & 1.026 & 1.024 & 1.002 & 1.020 \\ \hline
\textbf{17:00} & 1.050 & 1.050 & 1.048 & 1.046 & 1.043 & 1.041 & 1.081 \\ \hline
\textbf{18:00} & 1.088 & 1.085 & 1.091 & 1.085 & 1.080 & 1.084 & 1.124 \\ \hline
\textbf{19:00} & 1.116 & 1.113 & 1.121 & 1.116 & 1.103 & 1.112 & 1.153 \\ \hline
\textbf{20:00} & 1.117 & 1.112 & 1.117 & 1.116 & 1.097 & 1.105 & 1.154 \\ \hline
\textbf{21:00} & 1.106 & 1.103 & 1.102 & 1.100 & 1.089 & 1.097 & 1.144 \\ \hline
\textbf{22:00} & 1.099 & 1.097 & 1.092 & 1.091 & 1.082 & 1.098 & 1.134 \\ \hline
\textbf{23:00} & 1.047 & 1.045 & 1.038 & 1.038 & 1.040 & 1.062 & 1.077 \\ \hline
\end{tabular}}
\end{table}

\pagebreak

\begin{table}[h]
\centering
\caption{Mean difference of SIs for each hour and day from 2006–2017 ($\hat{\epsilon}^{d, h}$)}
\label{Table_SIAbsoluteDifference_BG}
\resizebox{\textwidth}{!}{%
\begin{tabular}{|c|c|c|c|c|c|c|c|}
\hline
\textbf{Time} & \multicolumn{1}{c|}{\textbf{Monday}} & \multicolumn{1}{c|}{\textbf{Tuesday}} & \multicolumn{1}{c|}{\textbf{Wednesday}} & \multicolumn{1}{c|}{\textbf{Thursday}} & \multicolumn{1}{c|}{\textbf{Friday}} & \multicolumn{1}{c|}{\textbf{Saturday}} & \multicolumn{1}{c|}{\textbf{Sunday}} \\ \hline
\textbf{00:00} & 0.174 & 0.166 & 0.174 & 0.168 & 0.170 & 0.193 & 0.186 \\ \hline
\textbf{01:00} & 0.107 & 0.097 & 0.104 & 0.098 & 0.100 & 0.120 & 0.123 \\ \hline
\textbf{02:00} & 0.064 & 0.057 & 0.059 & 0.059 & 0.060 & 0.075 & 0.080 \\ \hline
\textbf{03:00} & 0.112 & 0.113 & 0.106 & 0.115 & 0.107 & 0.062 & 0.057 \\ \hline
\textbf{04:00} & 0.205 & 0.206 & 0.197 & 0.208 & 0.196 & 0.107 & 0.076 \\ \hline
\textbf{05:00} & 0.257 & 0.258 & 0.249 & 0.260 & 0.245 & 0.181 & 0.156 \\ \hline
\textbf{06:00} & 0.228 & 0.223 & 0.216 & 0.226 & 0.218 & 0.217 & 0.209 \\ \hline
\textbf{07:00} & 0.150 & 0.134 & 0.130 & 0.132 & 0.134 & 0.179 & 0.195 \\ \hline
\textbf{08:00} & 0.077 & 0.064 & 0.063 & 0.062 & 0.065 & 0.115 & 0.143 \\ \hline
\textbf{09:00} & 0.053 & 0.049 & 0.048 & 0.046 & 0.048 & 0.065 & 0.091 \\ \hline
\textbf{10:00} & 0.050 & 0.048 & 0.048 & 0.050 & 0.053 & 0.058 & 0.057 \\ \hline
\textbf{11:00} & 0.050 & 0.048 & 0.050 & 0.055 & 0.056 & 0.056 & 0.058 \\ \hline
\textbf{12:00} & 0.049 & 0.047 & 0.050 & 0.051 & 0.047 & 0.060 & 0.066 \\ \hline
\textbf{13:00} & 0.049 & 0.054 & 0.048 & 0.054 & 0.050 & 0.047 & 0.044 \\ \hline
\textbf{14:00} & 0.043 & 0.048 & 0.045 & 0.049 & 0.046 & 0.076 & 0.097 \\ \hline
\textbf{15:00} & 0.080 & 0.088 & 0.090 & 0.085 & 0.087 & 0.137 & 0.164 \\ \hline
\textbf{16:00} & 0.138 & 0.146 & 0.146 & 0.144 & 0.139 & 0.172 & 0.191 \\ \hline
\textbf{17:00} & 0.141 & 0.142 & 0.140 & 0.141 & 0.131 & 0.143 & 0.160 \\ \hline
\textbf{18:00} & 0.086 & 0.081 & 0.078 & 0.081 & 0.081 & 0.083 & 0.088 \\ \hline
\textbf{19:00} & 0.122 & 0.120 & 0.122 & 0.119 & 0.123 & 0.121 & 0.125 \\ \hline
\textbf{20:00} & 0.173 & 0.172 & 0.168 & 0.161 & 0.156 & 0.152 & 0.178 \\ \hline
\textbf{21:00} & 0.238 & 0.236 & 0.234 & 0.227 & 0.210 & 0.204 & 0.239 \\ \hline
\textbf{22:00} & 0.284 & 0.278 & 0.279 & 0.271 & 0.254 & 0.246 & 0.285 \\ \hline
\textbf{23:00} & 0.252 & 0.250 & 0.247 & 0.245 & 0.241 & 0.244 & 0.262 \\ \hline
\end{tabular}}
\end{table}

\pagebreak

\begin{table}[h]
\centering
\caption{Mean difference as a percentage of Mean SI ($\delta^{d, h}$)}
\label{Table_SIPercentageDifference_BG}
\resizebox{\textwidth}{!}{%
\begin{tabular}{|c|c|c|c|c|c|c|c|c|c|c|}
\hline
\textbf{Time} & \multicolumn{1}{c|}{\textbf{Monday}} & \multicolumn{1}{c|}{\textbf{Tuesday}} & \multicolumn{1}{c|}{\textbf{Wednesday}} & \multicolumn{1}{c|}{\textbf{Thursday}} & \multicolumn{1}{c|}{\textbf{Friday}} & \multicolumn{1}{c|}{\textbf{Saturday}} & \multicolumn{1}{c|}{\textbf{Sunday}} & \multicolumn{1}{c|}{\textbf{Weekday}} & \multicolumn{1}{c|}{\textbf{Weekend}} & \multicolumn{1}{c|}{\textbf{Week}} \\ \hline
\textbf{00:00} & 18.74 & 17.61 & 18.31 & 17.73 & 17.75 & 19.01 & 18.66 & 18.03 & 18.84 & 18.26 \\ \hline
\textbf{01:00} & 12.47 & 11.08 & 11.80 & 11.15 & 11.27 & 12.78 & 13.29 & 11.56 & 13.04 & 11.98 \\ \hline
\textbf{02:00} & 7.76 & 6.77 & 6.96 & 6.98 & 7.03 & 8.34 & 9.04 & 7.10 & 8.69 & 7.55 \\ \hline
\textbf{03:00} & 13.87 & 13.69 & 12.76 & 13.83 & 12.78 & 7.06 & 6.63 & 13.39 & 6.85 & 11.52 \\ \hline
\textbf{04:00} & 25.49 & 25.03 & 23.89 & 25.22 & 23.53 & 12.34 & 8.97 & 24.63 & 10.65 & 20.64 \\ \hline
\textbf{05:00} & 30.91 & 30.56 & 29.38 & 30.69 & 28.75 & 20.75 & 18.33 & 30.06 & 19.54 & 27.05 \\ \hline
\textbf{06:00} & 25.24 & 24.52 & 23.61 & 24.80 & 23.74 & 24.31 & 24.27 & 24.38 & 24.29 & 24.35 \\ \hline
\textbf{07:00} & 15.31 & 13.70 & 13.25 & 13.43 & 13.61 & 19.18 & 22.04 & 13.86 & 20.61 & 15.79 \\ \hline
\textbf{08:00} & 7.49 & 6.23 & 6.13 & 6.03 & 6.31 & 11.72 & 15.32 & 6.44 & 13.52 & 8.46 \\ \hline
\textbf{09:00} & 5.00 & 4.66 & 4.58 & 4.37 & 4.55 & 6.38 & 9.23 & 4.63 & 7.81 & 5.54 \\ \hline
\textbf{10:00} & 4.73 & 4.58 & 4.60 & 4.78 & 5.06 & 5.63 & 5.63 & 4.75 & 5.63 & 5.00 \\ \hline
\textbf{11:00} & 4.72 & 4.59 & 4.80 & 5.27 & 5.34 & 5.41 & 5.63 & 4.95 & 5.52 & 5.11 \\ \hline
\textbf{12:00} & 4.65 & 4.52 & 4.84 & 4.92 & 4.51 & 5.80 & 6.37 & 4.68 & 6.09 & 5.09 \\ \hline
\textbf{13:00} & 4.64 & 5.17 & 4.62 & 5.17 & 4.79 & 4.61 & 4.32 & 4.88 & 4.46 & 4.76 \\ \hline
\textbf{14:00} & 4.11 & 4.65 & 4.38 & 4.75 & 4.48 & 7.63 & 9.76 & 4.47 & 8.70 & 5.68 \\ \hline
\textbf{15:00} & 7.69 & 8.51 & 8.77 & 8.26 & 8.50 & 13.89 & 16.51 & 8.35 & 15.20 & 10.30 \\ \hline
\textbf{16:00} & 13.30 & 14.16 & 14.22 & 14.03 & 13.58 & 17.17 & 18.73 & 13.86 & 17.95 & 15.03 \\ \hline
\textbf{17:00} & 13.43 & 13.53 & 13.36 & 13.48 & 12.56 & 13.73 & 14.80 & 13.27 & 14.27 & 13.56 \\ \hline
\textbf{18:00} & 7.91 & 7.46 & 7.15 & 7.46 & 7.50 & 7.65 & 7.83 & 7.50 & 7.74 & 7.57 \\ \hline
\textbf{19:00} & 10.93 & 10.78 & 10.88 & 10.67 & 11.15 & 10.88 & 10.84 & 10.88 & 10.86 & 10.88 \\ \hline
\textbf{20:00} & 15.48 & 15.47 & 15.05 & 14.43 & 14.22 & 13.76 & 15.42 & 14.93 & 14.59 & 14.83 \\ \hline
\textbf{21:00} & 21.53 & 21.39 & 21.23 & 20.64 & 19.28 & 18.59 & 20.89 & 20.81 & 19.74 & 20.51 \\ \hline
\textbf{22:00} & 25.83 & 25.35 & 25.54 & 24.84 & 23.48 & 22.40 & 25.14 & 25.01 & 23.77 & 24.65 \\ \hline
\textbf{23:00} & 24.06 & 23.93 & 23.79 & 23.59 & 23.16 & 22.98 & 24.34 & 23.71 & 23.66 & 23.69 \\ \hline
\midrule
\textbf{Overall} & 13.55 & 13.25 & 13.08 & 13.19 & 12.79 & 13.00 & 13.83 & 13.17 & 13.42 & 13.24 \\ \hline
\end{tabular}}
\end{table}

\pagebreak
\begin{figure}[H]
		\caption{Box plots for Seasonality indices (Bulgaria)}
		\begin{subfigure}[b]{0.4\textwidth}
			\includegraphics[width=\textwidth]{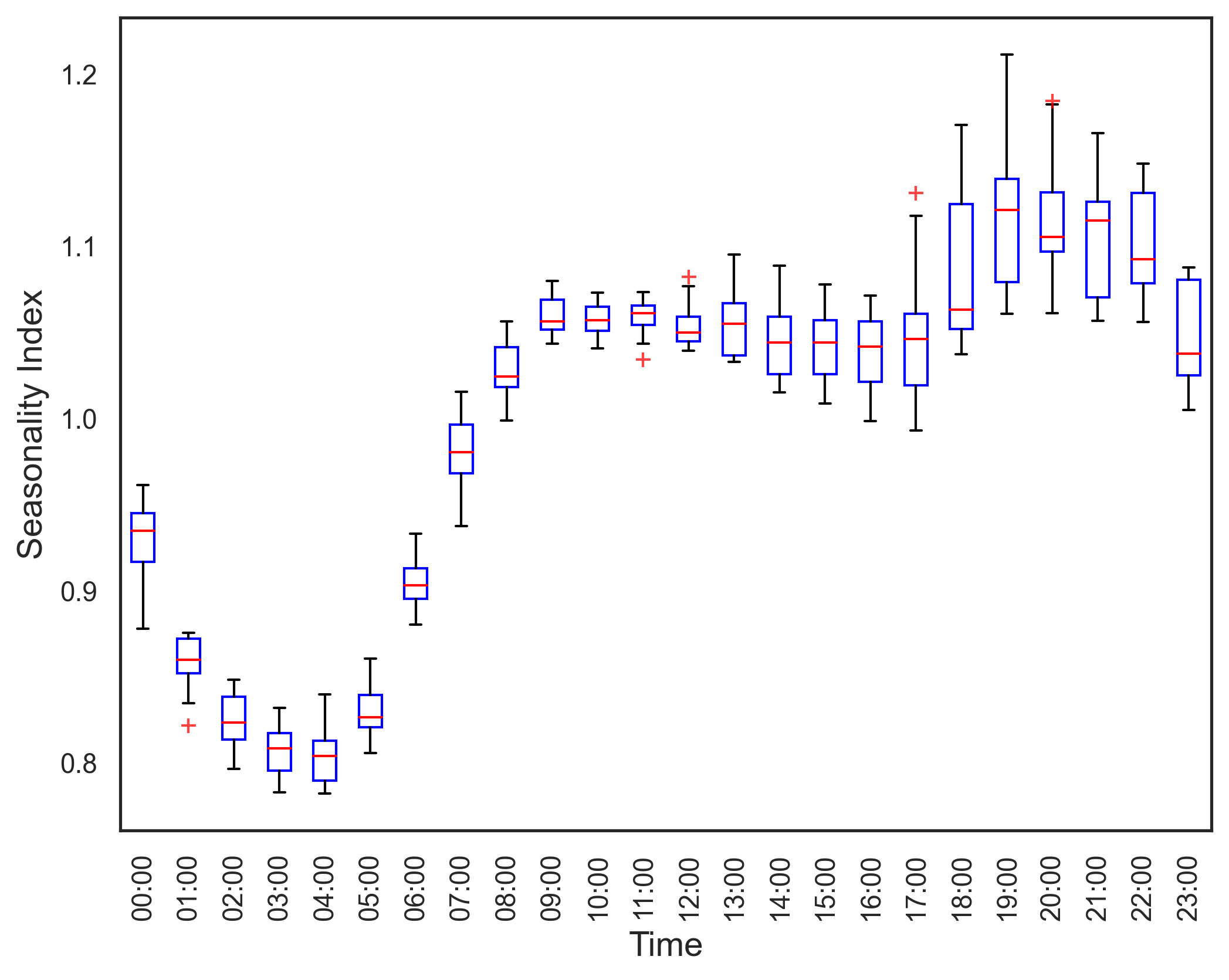}
			\caption{Monday}
			\label{Fig_Boxplots_SIMondayBG}
		\end{subfigure}
		\hfill
		\begin{subfigure}[b]{0.4\textwidth}
			\includegraphics[width=\textwidth]{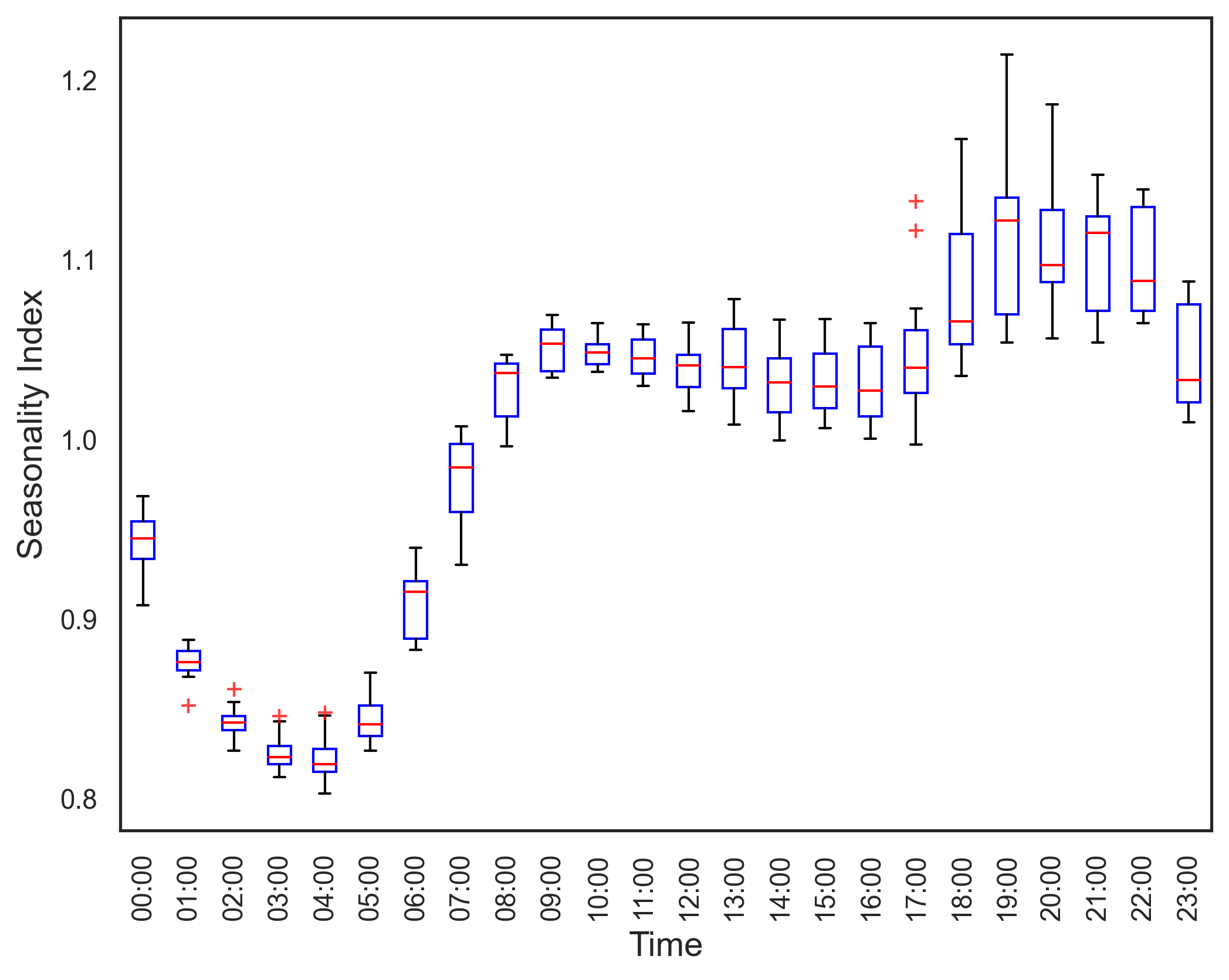}
			\caption{Tuesday}
			\label{Fig_Boxplots_SITuesdayBG}
		\end{subfigure}
		\hfill
  \begin{subfigure}[b]{0.4\textwidth}
			\includegraphics[width=\textwidth]{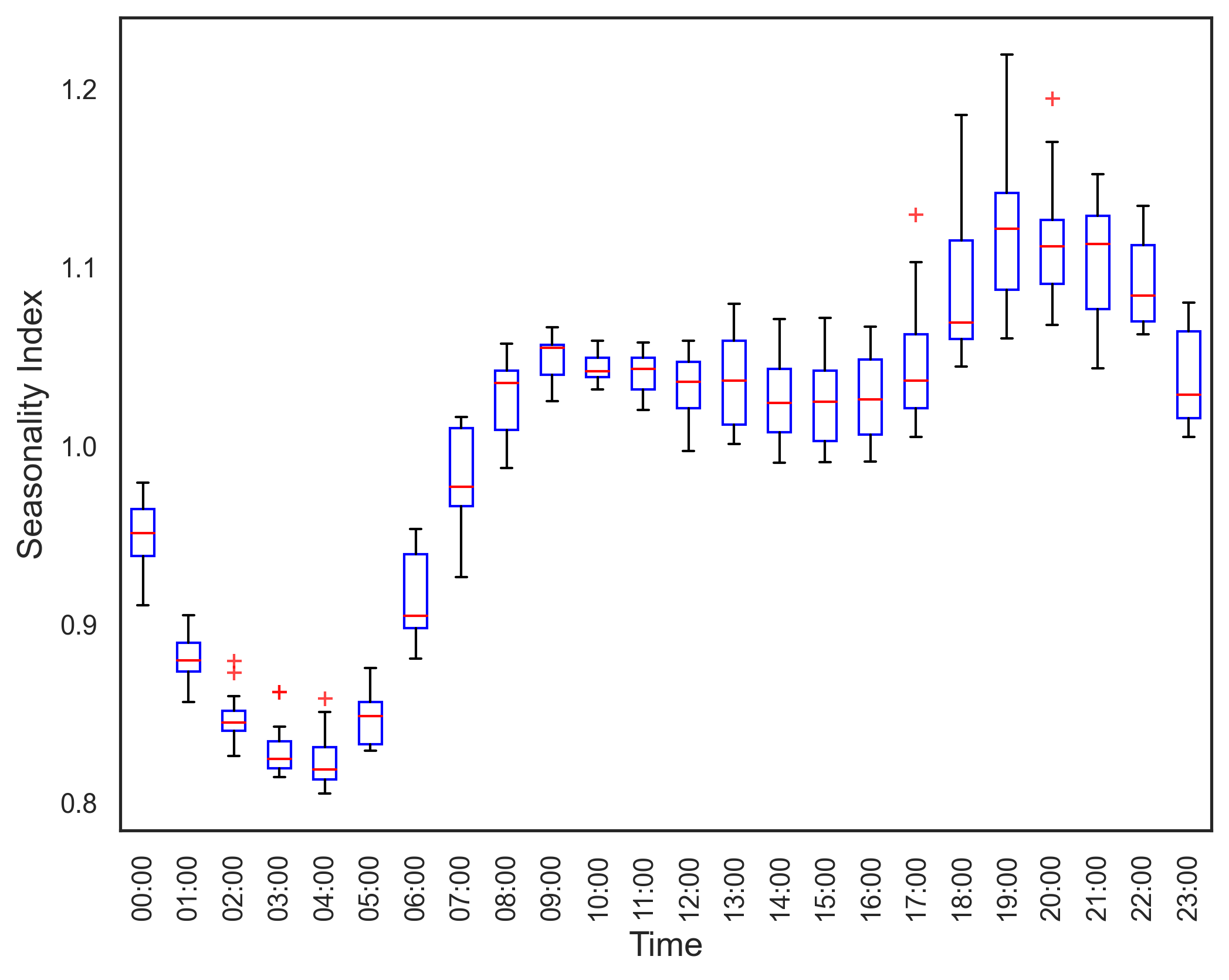}
			\caption{Wednesday}
			\label{Fig_Boxplots_SIWednesdayBG}
		\end{subfigure}
		\hfill
  \begin{subfigure}[b]{0.4\textwidth}
			\includegraphics[width=\textwidth]{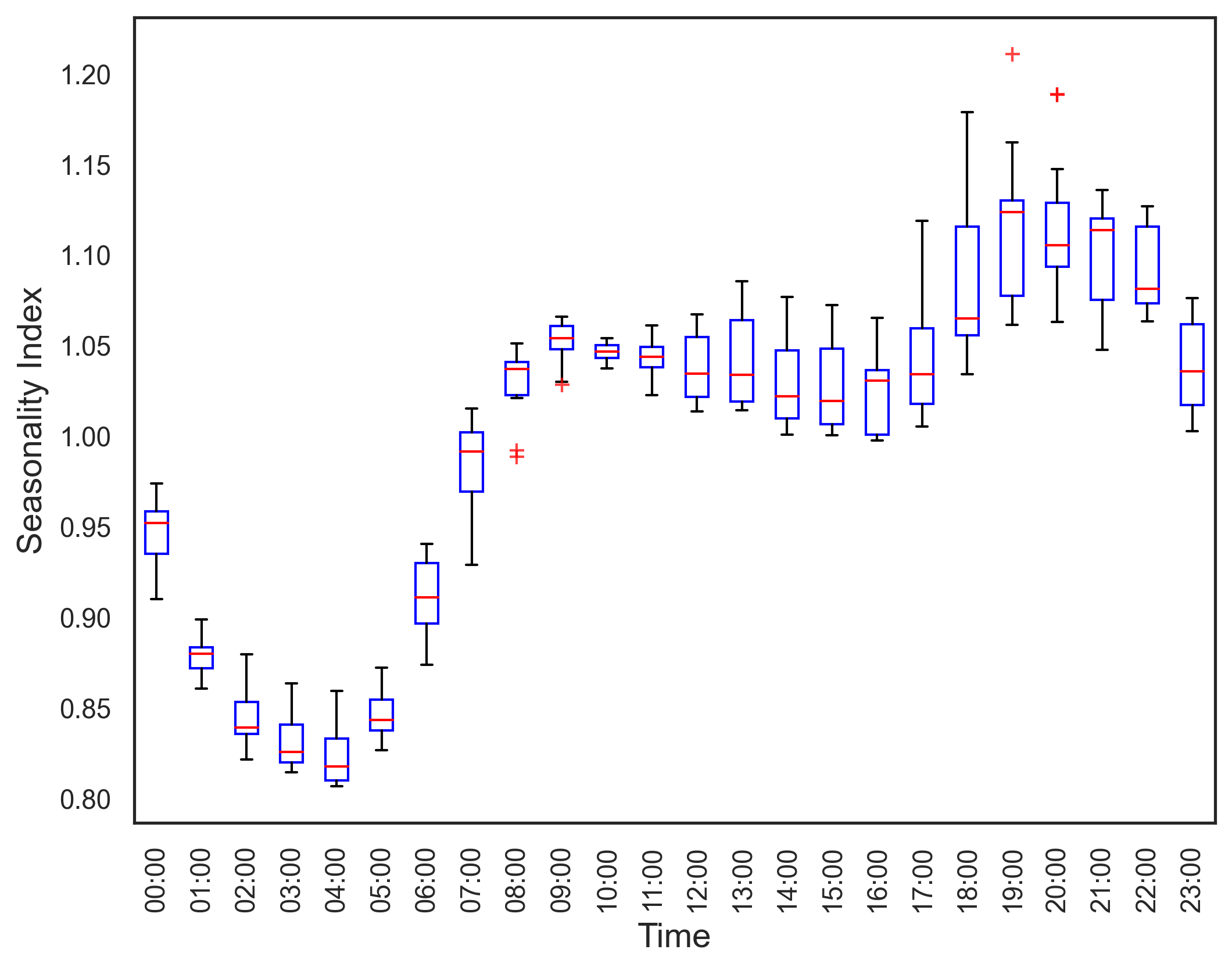}
			\caption{Thursday}
			\label{Fig_Boxplots_SIThursdayBG}
		\end{subfigure}
		\hfill
  \begin{subfigure}[b]{0.4\textwidth}
			\includegraphics[width=\textwidth]{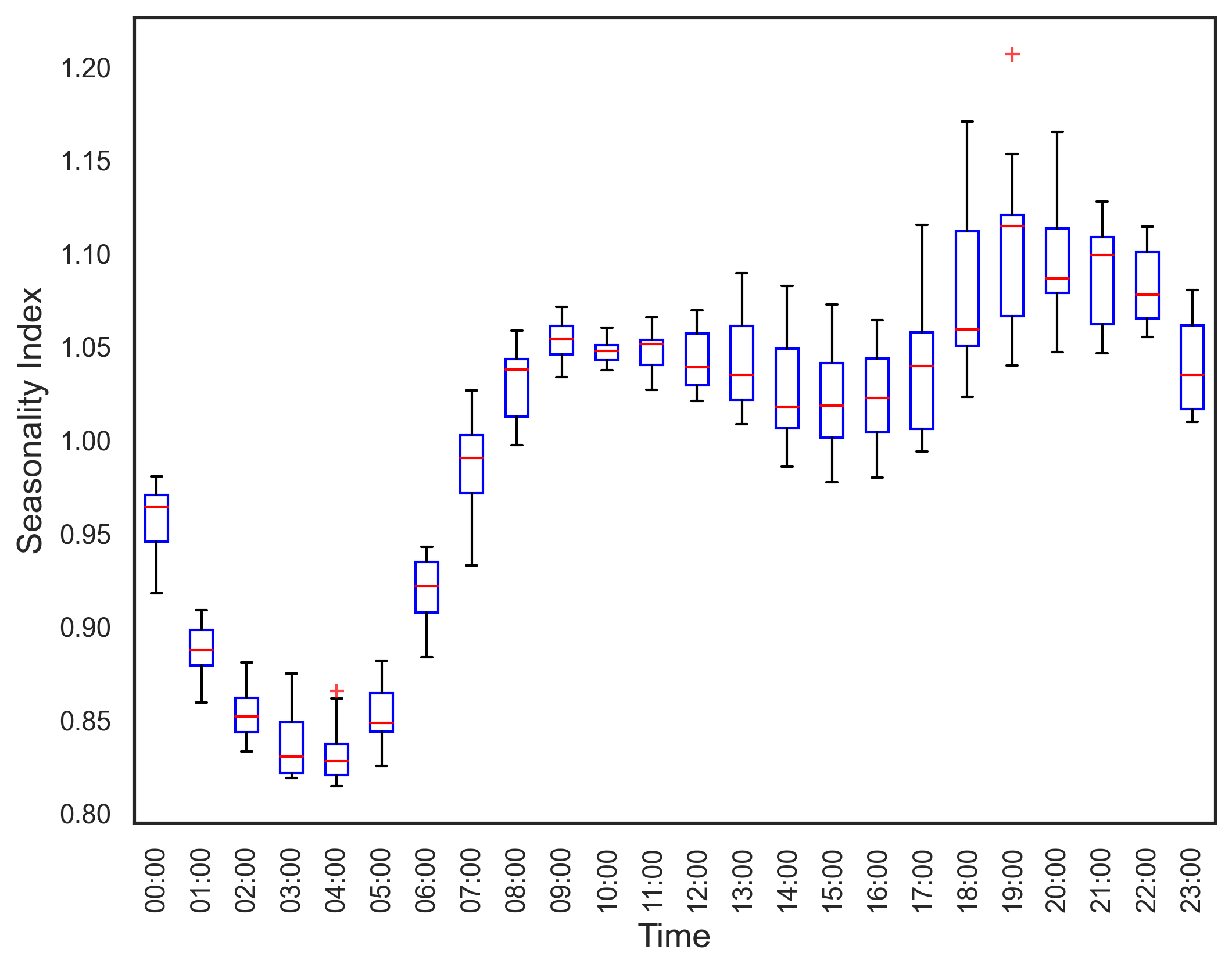}
			\caption{Friday}
			\label{Fig_Boxplots_SIFridayBG}
		\end{subfigure}
		\hfill
  \begin{subfigure}[b]{0.4\textwidth}
			\includegraphics[width=\textwidth]{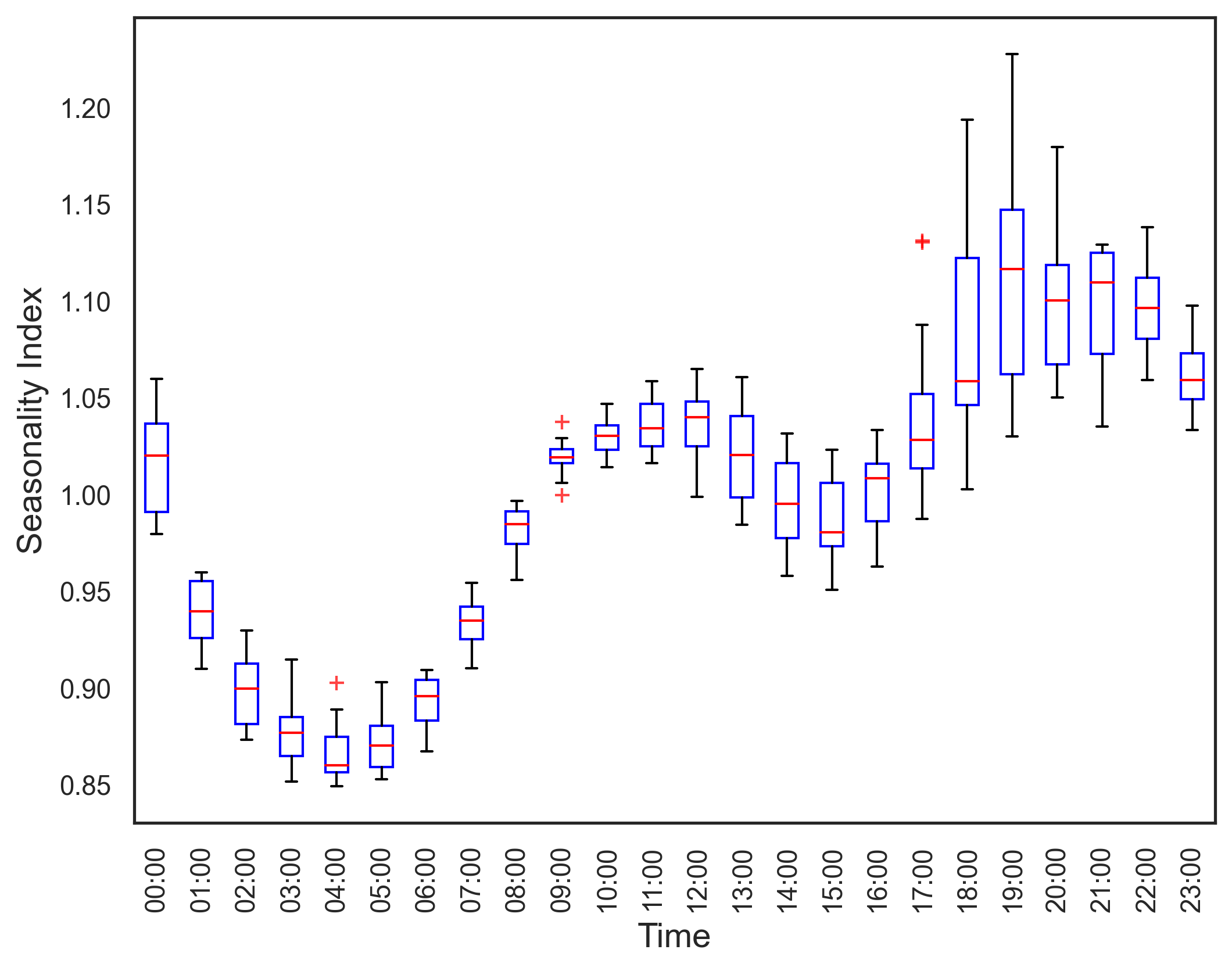}
			\caption{Saturday}
			\label{Fig_Boxplots_SISaturdayBG}
		\end{subfigure}
		\hfill
  \begin{subfigure}[b]{0.4\textwidth}
			\includegraphics[width=\textwidth]{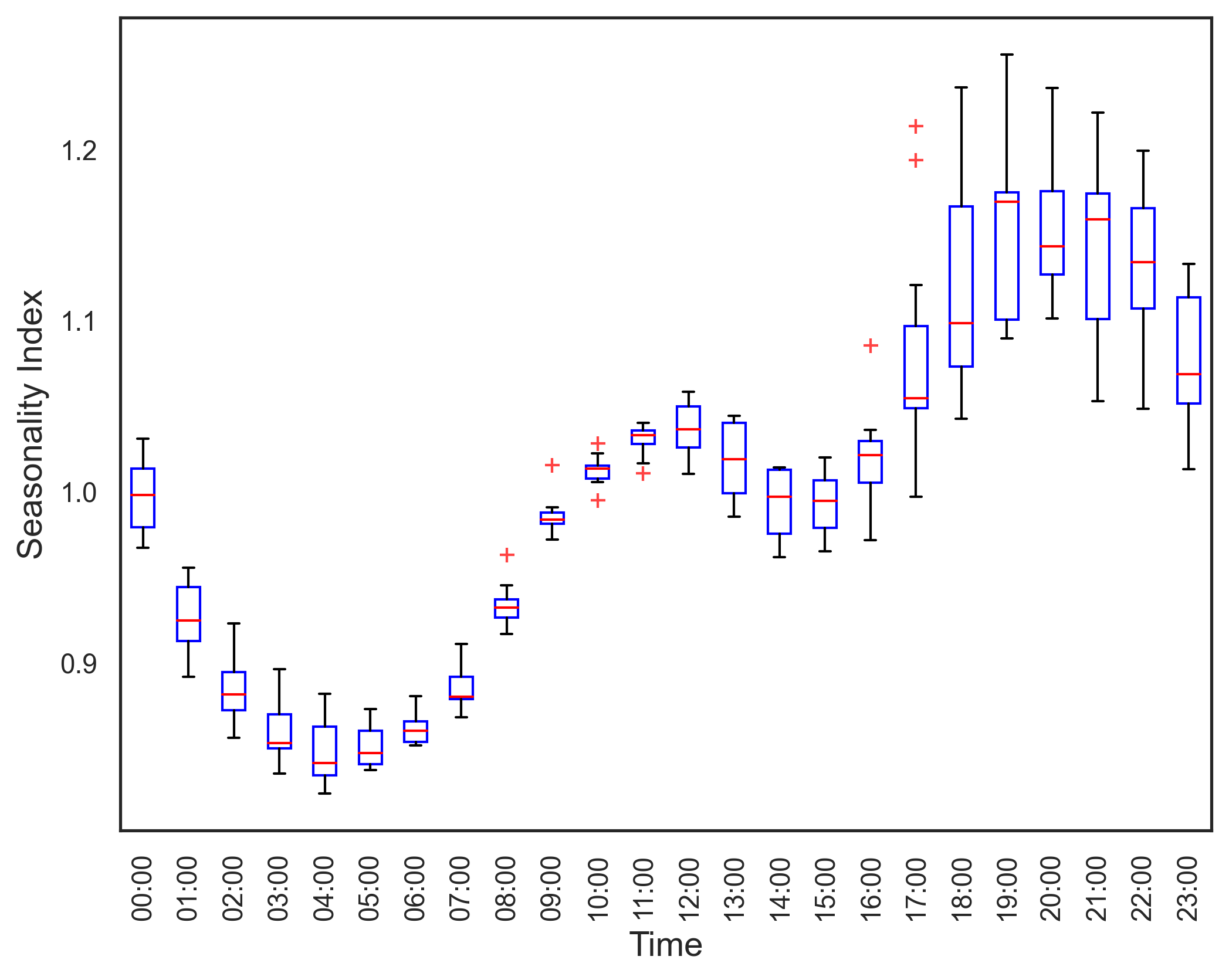}
			\caption{Sunday}
			\label{Fig_Boxplots_SISundayBG}
		\end{subfigure}
		\hfill
	\end{figure}

\pagebreak

\section{Descriptions of Alternative Short-term Forecasting Methods}\label{Appendix_ML}

\subsection{FBProphet:}

The first machine learning model we used is FBProphet developed by Meta. FBProphet employs an additive model that includes components for trend, seasonality (modeled using Fourier series), holidays, and potential changepoints. The trend is modeled as a piece-wise linear or logistic function, while seasonality is captured by Fourier series with user-defined periodicity. It incorporates Bayesian methods for modeling uncertainty and allows for the inclusion of custom seasonalities and prior knowledge (\citeauthor{taylor_prophet_paper_2017}, \citeyear{taylor_prophet_paper_2017}). This flexibility makes it suitable for various industries where irregular patterns are common, enabling efficient demand forecasting in retail, e-commerce, and social media analytics (\citeauthor{taylor_prophet_blog_2017}, \citeyear{taylor_prophet_blog_2017}). We implemented FBProphet in Python using the Prophet library.


\subsection{SARIMAX:}

Second, we looked at Seasonal Autoregressive Integrated Moving Average with Exogenous Variables (SARIMAX). SARIMAX is a time series forecasting model combining autoregressive (AR), moving average (MA), seasonal (S), and exogenous variable (X) components (\citeauthor{korstanje_sarimax_2021}, \citeyear{korstanje_sarimax_2021}). The ARIMA (AutoRegressive Integrated Moving Average) part models the temporal structure, while the exogenous variables extend the model's applicability beyond purely time-based patterns. It optimizes parameters through maximum likelihood estimation making it adaptable for various industries including stock price forecasting in finance, demand forecasting in economics, and weather patterns in meteorology. We implemented SARIMAX in Python using the statsmodels library.



\subsection{MSTL+autoARIMA:}

MSTL+autoARIMA integrates Multiple Seasonal and Trend Decomposition using Loess (MSTL) with automatic ARIMA modeling. MSTL decomposes a time series into trend, seasonal, and remainder components using locally weighted regression (\citeauthor{bandara_mstl_2021}, \citeyear{bandara_mstl_2021}). Subsequently, automatic ARIMA identifies the best-fitting ARIMA model for the remainder series. This hybrid approach handles multiple seasonal patterns and complex variations, making it suitable for retail sales, stock market analysis, and industries with diverse seasonal trends. We implemented MSTL+autoARIMA in Python using the statsforecast library developed by Nixtla (see https://nixtlaverse.nixtla.io/).

\subsection{Dynamic Harmonic Regression:}

Dynamic Harmonic Regression employs harmonic functions to model seasonal and trend components dynamically. It captures changes in the frequency and amplitude of seasonal patterns using Fourier series (\citeauthor{hyndman_2018}, \citeyear{hyndman_2018}; \citeauthor{young_dynamic_1999}, \citeyear{young_dynamic_1999}). This method flexibly adapts to the periodicity present in the data, making it applicable in time series forecasting, meteorology, and economics. Though less commonly used in electricity demand forecasting, its strength lies in its ability to handle periodic fluctuations in various datasets. We implemented Dynamic Harmonic Regression in Python using the Prophet library developed by Meta.

\subsection{GDP and Population Regression Models}

We also use GDP and Population Regression Models to compare with our ES model. We obtain the total yearly demand for a specific year using GDP and Population as predictors. Next, we divide the total yearly demand proportionally given the seasonality indices for each day and hour. These values would then serve as the forecasts for each hour.

\section{OECD and Non-OECD countries}
The OECD (Organization for Economic Cooperation and Development) is a group of 38 mostly high-income, democratic countries with market-based economies, while non-OECD countries encompass the rest of the world, including developing and emerging economies. As stated on their own website: "The OECD provides a forum in which governments can work together to share experiences and seek solutions to common problems. We work with governments to understand what drives economic, social and environmental change. We measure productivity and global flows of trade and investment. We analyse and compare data to predict future trends. We set international standards on a wide range of things, from agriculture and tax to the safety of chemicals." Two major differences between OECD countries and non-OECD countries are the amount of primary energy that they consume and their population growth. The countries that participate in OECD tend to be wealthier countries and use quite a bit more primary energy per capita

\end{document}